\documentclass[useAMS,usenatbib]{mn2e}

\usepackage{graphicx}
\usepackage{times}%
\usepackage{natbib}
\usepackage{graphics}
\usepackage{epsfig}
\usepackage{array}
\usepackage{stfloats}
\usepackage{fixltx2e}
\usepackage{amsmath}

\newcommand{\lesssim}{\la} 

\newcommand{\ie}{{i.e.}}
\newcommand{\eg}{{e.g.}}
\newcommand{\gsim}{\,\lower2truept\hbox{${>\atop\hbox{\raise4truept\hbox{$\sim$}}}$}\,}
\newcommand{\pp}{~~~.}
\newcommand{\vv}{~~~,}

\def\eg{{\rm e.g.$\,$}}
\def\ie{{\rm i.e.$\,$}}

\newcommand{\be}{\begin{equation}}
\newcommand{\ee}{\end{equation}}
\newcommand{\bea}{\begin{eqnarray}}
\newcommand{\eea}{\end{eqnarray}}

\renewcommand{\vec}[1]{ {\bmath #1} } 

\def\ltsima{$\; \buildrel < \over \sim \;$}
\def\simlt{\lower.5ex\hbox{\ltsima}}
\def\gtsima{$\; \buildrel > \over \sim \;$}
\def\simgt{\lower.5ex\hbox{\gtsima}}

\title[Oscillating structures in growing neutrino quintessence]{Oscillating nonlinear large scale structure in growing neutrino quintessence}

\author[M. Baldi et al.]{Marco Baldi$^{1,2}$, Valeria Pettorino$^{3}$, Luca Amendola$^{4}$, Christof Wetterich$^{4}$
\\$^{1}$~Excellence Cluster Universe, Boltzmannstr.~2, D-85748 Garching, Germany
\\$^{2}$~University Observatory, Ludwig-Maximillians University Munich, Scheinerstr. 1, D-81679 Munich, Germany
\\$^{3}$~SISSA, Via Bonomea 265, 34136 Trieste, Italy
\\$^{4}$~Institute for Theoretical Physics, Heidelberg University, Philosophenweg 16, D-69120 Heidelberg, Germany}
\begin{document}


\pagerange{\pageref{firstpage}--\pageref{lastpage}} \pubyear{2011}

\maketitle

\label{firstpage}

\begin{abstract}
\ \\	
Growing Neutrino quintessence describes a form of dynamical dark energy that
could explain why dark energy dominates the universe only in recent cosmological
times. This scenario predicts the formation of large scale neutrino lumps which
could allow for observational tests. We perform for the first time N-body
simulations of the nonlinear growth of structures for cold dark matter and
neutrino fluids in the context of Growing Neutrino cosmologies. 
Our analysis shows a pulsation - increase and subsequent decrease - of the
neutrino density contrast. This could lead to interesting observational
signatures, as an enhanced bulk flow in a situation where the dark matter
density contrast only differs very mildly from the standard $\Lambda$CDM
scenario. We also determine for the first time the statistical distribution of
neutrino lumps as a function of mass at different redshifts. Such determination
provides an essential ingredient for a realistic estimate of the observational
signatures of Growing Neutrino cosmologies. Due to a breakdown of the 
non-relativistic Newtonian approximation our results are limited to redshifts $z\ge 1$.

\end{abstract}



\section{Introduction}
\label{i}

The presently accepted cosmological model describes the evolution of the Universe by assuming the existence of two distinct forms of gravitating energy
that largely dominate the total cosmic density, leaving to the particles described by the standard model of particle physics only a small fraction
of the energy budget. These two components are a family of non-relativistic massive particles (Cold Dark Matter, CDM hereafter) that source the
gravitational instability processes leading to the formation of galaxies and galaxy clusters, and a diffuse component with a negative equation 
of state (Dark Energy, DE hereafter) that
accounts for the observed accelerated expansion \citep{Riess_etal_1998, Perlmutter_etal_1999,Kowalski_etal_2008} of the Universe 
during the last six billion years of its evolution.

For the latter component, the standard model assumes a cosmological constant $\Lambda $, \ie a perfectly homogeneous form of energy whose density
remains constant in time despite the expansion of the Universe. Although this simple choice reproduces with remarkable accuracy a large amount of observational data, it is faced by two fundamental theoretical problems concerning its fine-tuned value (known as the ``fine-tuning problem") and the
coincidence of its domination over cold dark matter (CDM) only at
a relatively recent cosmological epochs (known as the ``why now?" problem).
In order to provide possible solutions to these issues, alternative scenarios for the cosmic DE have been proposed. In particular, 
a viable alternative consists in identifying the DE with a scalar field $\phi $ (the ``cosmon") dynamically evolving in a self-interaction potential $V(\phi )$, 
as proposed by \cite{Wetterich_1988} and \citet{Ratra_Peebles_1988}.

For the former component, instead, several candidates can be considered \citep[see][for a review]{Bertone_Hooper_Silk_2005}, including weakly interactive
massive particles (WIMPS) as the neutralino \citep{Goldberg_1983,Ellis_etal_1984}, arising in supersymmetric extensions of the standard model, 
or light scalars as the axion \citep{Preskill_Wise_Wilczek_1983, Rosenberg_VanBibber_2000} or geon \citep{Beyer_etal_2010}. Massive neutrinos, being an already known family of massive particles beyond the standard model
would seem the most natural candidate for dark matter. They are, however, ruled out as the main contributor to the dark matter density 
by observational constraints on the evolution of structure formation processes \citep[see \eg ][]{Bond_Szalay_1983}.

Nevertheless, even a small fraction of dark matter in the form of massive neutrinos could play a crucial role in the evolution of the Universe and
in the late stages of large scale structure formation, in particular if a direct interaction between massive neutrinos and a DE scalar field is present, 
as will be discussed in the present paper.
\ \\

Cosmologies with an increasing neutrino mass stopping the evoltuion of dark energy due to
a coupling between neutrinos and a DE scalar field, known as the ``Growing Neutrino" scenario, 
have been first proposed by \citet{Amendola_Baldi_Wetterich_2008, Wetterich_2007} with the aim
to provide a viable explanation of the ``why now?" problem. 
In such models, in fact, the reason behind the present quasi-static behavior of the DE density and equation of state resides in the direct cosmon-neutrino coupling. 
As a consequence of the interaction, the neutrino mass grows in time depending on the value of the cosmon. 
However, like in standard coupled DE models \citep{Wetterich_1995,Amendola_2000}, the interaction with relativistic species is suppressed for symmetry reasons.
This implies that the coupling between massive neutrinos and the DE scalar field gets active only when neutrinos become non-relativistic.

In the context of Growing Neutrino models, therefore, the neutrino mass may be very small at decoupling, and grow into the eV range only in a rather recent cosmological epoch.
A transition between relativistic and non-relativistic neutrinos will then occur only at relatively recent redshifts, $z \sim 5-10$. 
At higher redshifts neutrinos will free-stream and behave as relativistic particles.
At $z \lesssim 5$, the time evolution of the neutrino mass changes the evolution of the cosmon itself via the Klein Gordon equation:
\be \label{kg} \phi'' + 2{\cal H} \phi' + a^2 \frac{dV}{d \phi} = a^2 \beta
(\rho_{\nu}-3 p_{\nu}) \,\, , \ee with $\rho_\nu$ and $p_\nu = w_\nu \rho_\nu$ 
the energy density and pressure of the neutrinos.

As soon as this happens, the coupling to the neutrino fluid effectively stops the further time evolution of the cosmon field, and from this time on DE 
effectively behaves as a cosmological constant. 
In other words, when neutrinos become non-relativistic, the scalar field potential $V(\phi )$ acquires a correction due to the neutrino coupling and the scalar field
feels an effective potential with a slowly evolving minimum, where the DE scalar field stops its evolution.
Since the effective potential $V_{eff}$ differs from the true potential $V(\phi)$ only when neutrinos become nearly pressureless, Growing Neutrino models account 
for the sudden recent dark energy domination by relating it to a `cosmological event', i.e. neutrinos becoming massive enough to behave as a non-relativistic fluid.

When this transition happens, the cosmon field moves away from its early scaling solution. Subsequently, it almost stops in the minimum of the effective potential.
More precisely, the cosmon does not stop but oscillates around the minimum of the effective potential, leaving a characteristic oscillatory imprint also on the neutrino mass. In this paper we will describe for the first time in detail the effect that such an oscillating pattern leaves on structure formation.

The cosmon-neutrino coupling $\beta$ can be large \citep[as compared to gravitational strength, see \eg][]{Fardon_Nelson_Weiner_2004,Bjaelde_etal_2008,Brookfield_VanDeBruck_Mota_2006}, such that even the small fraction of energy density in cosmic neutrinos can have important cosmological effects. The fifth force mediated by the cosmon is substantially stronger than the standard gravitational interaction. Its effect becomes important only once the neutrino mass is sufficiently big for neutrinos to be non-relativistic. When this happens, neutrinos feel the presence of the fifth force and can collapse into stable bounded structures of the type described in \citet{Brouzakis_Tetradis_Wetterich_2008, Bernardini_Bertolami_2009}. 

It has been shown that neutrino lumps form at redshift $z_\text{nl} \approx 1-2$, when the neutrino fluctuations become nonlinear \citep{Mota_etal_2008, Wintergerst_etal_2010}. Furthermore, neutrino lumps significantly affect only large scales \citep{Mota_etal_2008, Pettorino_etal_2010} with typical sizes in the range of a few $10-100$ Mpc.

A correct evaluation of the size and time dependence of the gravitational
potential induced by neutrino lumps is essential (and yet difficult to achieve)
for a comparison of Growing Neutrino models with current available data.
Furthermore, in a Newtonian approximation the role of the gravitational
potential is taken by $\beta\,\delta\phi$, with $\delta\phi$ the local
fluctuation of the cosmon or quintessence field. While the gravitational
potential remains much smaller than one even for highly nonlinear neutrino
lumps, the cosmon induced  potential $\beta\,\delta\phi$ can reach values of
order one, signalling a breakdown of linear theory in the cosmon sector. The
difficulty for the computation of the neutrino-induced gravitational potential
comes from the fact that an extrapolation of the linear evolution violates
rapidly even the most extreme bound which would result if all neutrinos of the
visible universe were concentrated in one spot \citep{Wintergerst_etal_2010}. In
order to get a meaningful picture of Growing Neutrino quintessence a statistical
analysis of the neutrino lump abundances as a function of their mass is
required. In order to achieve such a statistical understanding of neutrino
lumps, it is necessary to rely on non-linear methods, such as N-body
simulations, that can allow neutrinos to distribute, merge and diffuse with time
and within different lumps. 

Both the size and the time evolution of the neutrino-induced gravitational
potential are important features that need to be understood in order to draw a
realistic picture of the Growing Neutrino scenario. 
It was shown in \cite{Pettorino_etal_2010} that the lump potential can have
important effects on the propagation of photons and therefore influence the
CMB-anisotropies via the integrated Sachs-Wolfe (ISW) effect, which may be
observed by a modification of the CMB-spectrum at low angular multipoles. The
neutrino gravitational potential also influences the correlation between matter
and photon fluctuations for which the present observations yield values larger
than expected in the cosmological $\Lambda$CDM model \citep{Jain_Ralston_1999,
Inoue_Silk_2006, Rudnick_Brown_Williams_2007, Samal_Saha_Jain_2008,
Giannantonio_etal_2008}. Moreover, a sudden increase in $\Phi_\nu$ could
reconcile \citep{Ayaita_etal_2009} the presently observed large bulk flows
\citep{Watkins_etal_2009} with observational bounds on the matter fluctuations
at similar scales \citep{Reid_etal_2010}. The size of the gravitational
potential also determines the counter-effect of neutrino lumps onto dark matter
structures, possibly erasing baryonic oscillations from the CDM power spectrum
if the neutrino-potential exceeds substantially a value around $10^{-5}$
\citep{Brouzakis_etal_2011}. 

As already mentioned above, the difficulty in providing a realistic estimate of
the neutrino gravitational potential $\Phi _{\nu }$ resides in the fact that
linear theory breaks down. A first estimate of $\Phi _{\nu }$ based on
extrapolations relying on a careful matching between linear equations for $k <
10^{-4}$ h/Mpc and nonlinear results for $k > 2\cdot10^{-2}$ h/Mpc was attempted
in \cite{Pettorino_etal_2010}, providing also a first attempt to address the
backreaction of small neutrino lumps onto larger-scale lumps. 

The evolution of isolated lumps was addressed within non-linear hydrodynamic
equations in \cite{Wintergerst_etal_2010}, showing how they mimic very large
cold dark matter structures, with a typical local gravitational potential $10^{-5}$
for a lump size of $10$ Mpc. Spherical collapse within Growing Neutrino
quintessence \citep{Wintergerst_pettorino_2010} first showed that the
extrapolated linear density contrast at collapse manifests a characteristic
oscillating pattern that reflects the evolution of the background neutrino mass
$m_\nu(\phi)$ as the cosmon $\phi$ moves around the minimum of the effective
potential. Spherical collapse cannot, however, take into account effects due to
velocity directionality, which, as shown by \cite{Baldi_etal_2010}, can be very
relevant in the presence of a fifth force and can only be addressed in the
context of N-body simulations.
\ \\

In this paper we present for the first time the results of N-body simulations for Growing Neutrino models. This allows us to follow not only the growth of neutrino perturbations in the nonlinear regime, but also, for the first time, to estimate the statistical abundance of neutrino lumps as a function of their final mass. N-body simulations further permit to follow the initial stages of the formation of neutrino lumps with a detail never reached before: CDM structures first form, as in the standard scenario; then, we directly show how neutrinos start to concentrate along CDM filaments, the latter seeding the subsequent growth of neutrino lumps. 

Important differences need to be taken into account with respect to CDM simulations. First, it is necessary to switch between a phase in which neutrinos are relativistic and feel no coupling to a phase in which neutrino fluctuations grow under the effect of the fifth force; this transition happens around $z \sim 4$ for the Growing Neutrino model presented in \cite{Mota_etal_2008} and investigated here. Second, the strength of the coupling can be order $10^3$ stronger than gravitational attraction, with a consequent fast growth of neutrino lumps which will deserve particular care, both for its impact on the computational cost of the simulations and, physically, for its consequences on particle velocities. Indeed, the latter eventually limit the redshift down to which it is possible to rely on N-body simulations, valid as long as velocities are small enough compared to the speed of light. We discuss this issue in detail for the first time and show that the formation of neutrino lumps follows an oscillating pattern, depending on the background oscillations and on the interplay of mass and velocity dependent terms. 

After a rapid overview of Growing Neutrino quintessence both at the background  and at the linear levels in Sects.~\ref{sec:gnc} and \ref{sec:nl}, we summarize what we already know about neutrino lumps in Sect.~\ref{sec:nl}. We then describe the set of N-body simulations carried out for the present work and discuss in Sect.~\ref{sec:nb} the regime of validity of our analysis, ultimately limited by neutrino velocities growing up to relativistic values. In section \ref{sec:results} we present our results: we are able to follow neutrino particles and their velocities. Neutrino lumps, first seeded by CDM filamentary structures, evolve in an oscillating manner, becoming more and less concentrated as time goes by. We provide a physical intuitive explanation for this behavior in Subsec.~\ref{subsec:oscillations} and we compute the halo abundance of neutrino lumps as a function of their mass in Subsec.~\ref{subsec:statistics}. In Subsec.~\ref{subsec:cdm} we consider effects on the CDM matter power spectrum and give an estimate of the power enhancement due to the presence of neutrino structures. Finally, in Sect.~\ref{concl} we draw our conclusions.

\section{Growing neutrino cosmologies} \label{sec:gnc}

Growing Neutrino models involve a coupling between the DE scalar field (cosmon) and massive neutrinos. 
At the background level, the expansion of the universe obeys the Friedmann equation:
\be \label{f1} {\cal H}^2 \equiv \left(\frac{a'}{a}\right)^2 = \frac{a^2}{3} \sum_\alpha \rho_\alpha \ee
where primes denote derivatives with respect to conformal time $\tau$. The sum is taken over all components $\alpha$ of the energy density in the universe, including CDM, DE, neutrinos, baryons and radiation. The time evolution of the energy density $\rho_\alpha$ for each species involves the equation of state $w_\alpha \equiv p_\alpha/\rho_\alpha$. We use units in which the reduced Planck mass $M_{{\rm Pl}}/8\pi $ is set to one.

A crucial ingredient in this model is the dependence of the average neutrino mass on the cosmon field $\phi$, as encoded in the dimensionless cosmon-neutrino coupling $\beta$, \be 
\label{beta_phi} \beta \equiv - \frac{d \ln{m_\nu}}{d \phi} \pp 
\ee 
We consider here a constant $\beta < 0$ such that for increasing $\phi$ the neutrino mass increases with time 
\be 
\label{eq:nu_mass} m_{\nu} \equiv m_{\nu}(t_{0}) e^{-{{\beta}} \phi} \vv 
\ee 
where $\bar{m}_{\nu}$ is a constant. 
For a given
cosmological model with a given time dependence of $\phi$, one can determine
the time dependence of the neutrino mass $m_\nu(t)$. For three degenerate
neutrinos the present average value of the neutrino mass $m_\nu(t_0)$ can be related
to the energy fraction in neutrinos ($h \approx 0.72$) \be \label{omeganu} \Omega_{\nu} (t_0) = \frac{3 m_\nu
  (t_0)}{94\, eV h^2} \,\, . \ee 

For the purpose of this paper we will consider the model described in \cite{Mota_etal_2008}, with a constant coupling $\beta = -52$. This choice corresponds to a present neutrino mass of about $2.1$ eV. In general, $\beta$ can be a function of $\phi$, as proposed in \citet{Wetterich_2007} within a particle physics model, modifying equation (\ref{eq:nu_mass}) but leading to similar effects. 
We refer to \cite{Mota_etal_2008} for further details on this model and recall here for convenience only its essential ingredients.

The cosmon exchange mediates an additional attractive force between neutrinos of strength $2\beta^2$. The case $\beta \sim 1$ corresponds to a strength comparable to gravity. The dynamics of the cosmon can be inferred from Eq.~(\ref{kg}), where we choose
an exponential potential \cite{Wetterich_1988} \cite{Ratra_Peebles_1988} \cite{Ferreira_Joyce_1998}:
\be \label{pot_def} V(\phi) \propto e^{- \alpha \phi} \pp \ee
The constant $\alpha$ is one of the free parameters of our model and determines the slope of the potential and thus the DE fraction at early times. Current bounds constrain it to be of the order $\alpha \sim 10$ or bigger \citep{Doran_etal_2007} and we assume $\alpha = 10$ in the present work. The naturalness of the exponential potential in the presence of quantum fluctuations and a cosmon-matter coupling has been discussed by \cite{Wetterich_2008}. 

The homogeneous energy density and pressure of the scalar field $\phi$ are defined
in the usual way as \be \label{phi_bkg} \rho_{\phi} = \frac{\phi'^2}{2 a^2} + V(\phi)  \vv \,\,\, p_{\phi} = \frac{\phi'^2}{2 a^2} - V(\phi)  \vv \,\,\, w_{\phi} = \frac{p_{\phi}}{\rho_{\phi}} \pp \ee
We choose the value of $\gamma = -\beta/\alpha$ such that
we obtain the correct present dark energy fraction according to the relation \citep{Amendola_Baldi_Wetterich_2008}:
\be
\Omega_h(t_0) = \frac{\gamma m_\nu(t_0)}{16 eV} \,\,\,\,\, , \,\,\,\,\,\, \gamma = -\frac{\beta}{\alpha} \pp
\ee 
Dark energy and growing neutrinos follow the coupled conservation equations: 
\bea \label{cons_phi} \rho_{\phi}' = -3 {\cal H} (1 + w_\phi) \rho_{\phi} +
\beta \phi' (1-3 w_{\nu}) \rho_{\nu} \vv \\
\label{cons_gr} \rho_{\nu}' = -3 {\cal H} (1 + w_{\nu}) \rho_{\nu} - \beta \phi' (1-3 w_{\nu}) \rho_{\nu}
\pp \eea The sum of the energy momentum tensors for neutrinos and DE is conserved,
 but not the separate parts. We neglect a possible cosmon coupling
to CDM, so that $\label{cons_cdm} \rho_c' = -3 {\cal H} \rho_c $.

\begin{figure}
\begin{center}
\begin{picture}(185,235)(20,0)
\put(-6,200){{$\rho$}}
\put(110,-6){{$1+z$}}
\includegraphics[width=85mm,angle=0.]{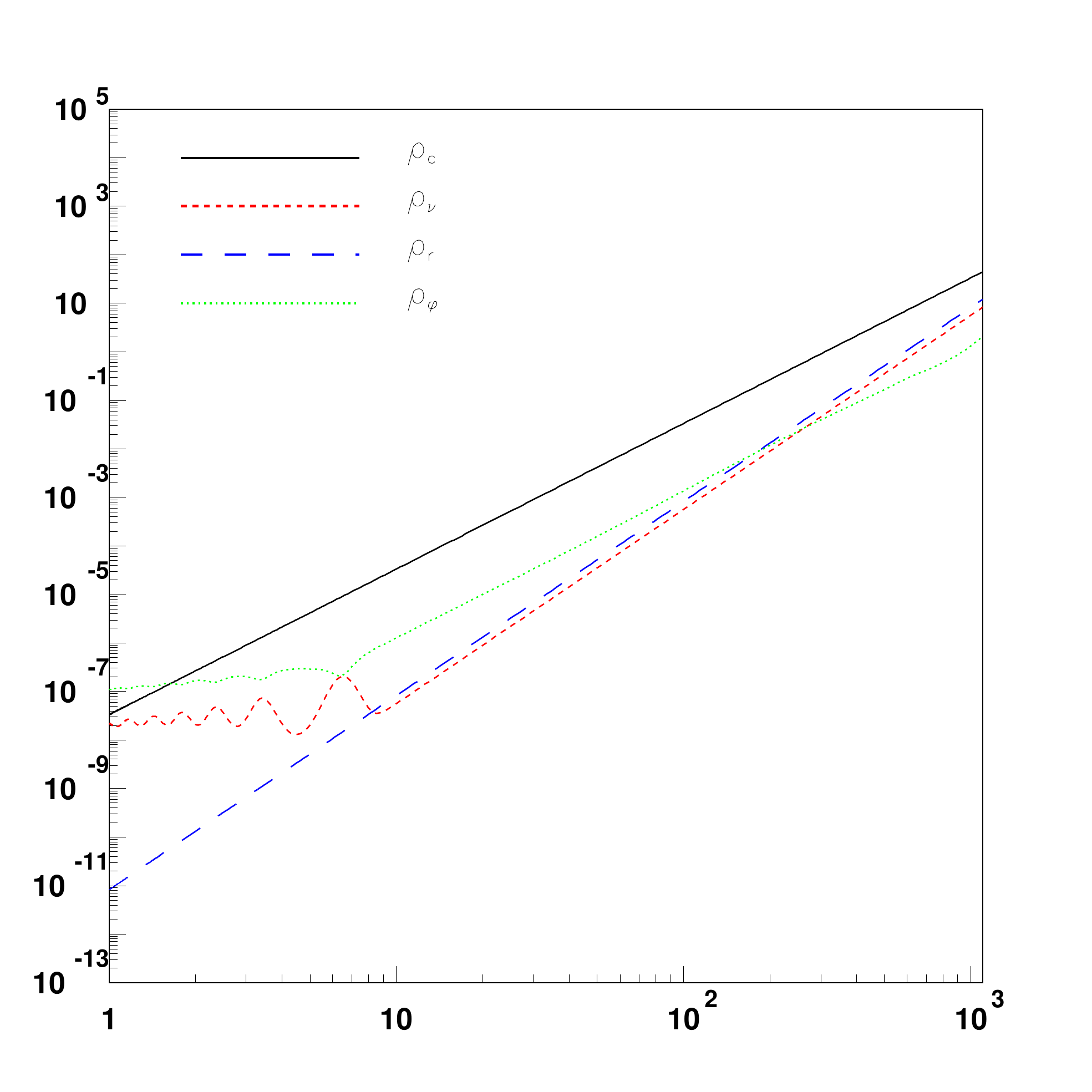}
\end{picture}
\end{center}
\caption{Energy densities of neutrinos (dashed, red), cold dark matter (solid, black), dark energy (dotted, green) and photons (long
dashed, blue) are plotted vs redshift. For all plots we take a constant $\beta = -52$, $\alpha = 10$ and present neutrino mass
$m_\nu = 2.11$ eV.}
\label{fig_1_const}
\vspace{0.5cm}
\end{figure}

Given the potential (\ref{pot_def}), the evolution equations for the different species can be numerically integrated, providing the background evolution shown in Fig.~\ref{fig_1_const} for a constant $\beta = -52$. The initial
pattern is a typical early dark energy model, since neutrinos are still relativistic and almost
massless, with $p_\nu = \rho_\nu/3$, so that the coupling term in eqs.(\ref{kg}), (\ref{cons_phi}), (\ref{cons_gr}) vanishes. DE
is still subdominant and follows the attractor solution provided by the
exponential potential \citep[see ][for details]{Wetterich_1995, Amendola_2000, Copeland_etal_1998, Mota_etal_2008}:
for $z \gsim 6$ it tracks the dominant background component with an early dark energy fraction $\Omega_h = n/\alpha^2$ and $n = 3(4)$ for the matter (radiation) dominated era. 
As soon as neutrinos become nonrelativistic, the coupling term $\sim \beta \rho_\nu$ in the evolution equation for the cosmon (Eq.~\ref{kg}) starts to play a significant role. This cosmological ``trigger event'' kicks $\phi$ out of the attractor. As shown in Fig.~{\ref{fig_1_const}}, $\rho_\nu$ and $\rho_\phi$ change behavior for $z < 6$: the value of the cosmon stays `almost' constant and the frozen cosmon potential mimics a cosmological constant. 
According to this model, at the present time neutrinos are still subdominant with respect to CDM, though in the future they will take the lead. For our choice of the parameters neutrino pressure terms may be safely neglected for redshifts $z_{nr}<4$; before that redshift neutrinos free stream as usual relativistic particles.

Most importantly,  the $\phi - \nu$ coupled fluid manifests small decaying oscillations: the cosmon moves around the minimum of the effective potential and $\rho_\nu$ oscillates towards an almost constant value. The cosmon background oscillations enter also in the evolution of the neutrino mass $m_\nu(\phi)$ which in turn influences the induced gravitational potential. Consequently, the extrapolated linear density at collapse oscillates in redshift, as shown in \cite{Wintergerst_pettorino_2010}.

In this paper we will address the consequences of such a pattern in great detail, showing for the first time how these oscillations directly affect structure formation. Before moving to the results of N-body simulations, however, we recall in the next section our present knowledge about large scale structures in Growing Neutrino models and, in particular, on ``neutrino lumps".

\section{Neutrino lumps} \label{sec:nl}

Linear perturbations in Growing Neutrino quintessence have been widely described in \cite{Mota_etal_2008} to which we refer for details. We recall here that, as it happens whenever a coupling of this type is active \citep{amendola_2004, pettorino_baccigalupi_2008, Baldi_etal_2010}, it is possible to reformulate the Euler equation in order to include the coupling contribution for each particle $\alpha$:
\begin{eqnarray} \label{euler}
& &{\bf{{v}_\alpha'}} + \left({\cal H} - {\beta} \phi' \right) {\bf
{v}_\alpha} - {\bf{\nabla }} \left[\Phi_\alpha + \beta \phi \right] = 0
\,.
\end{eqnarray}
In cosmic time (${\rm d}t = a\, {\rm d}\tau$) Eq.~(\ref{euler})  can also be
rewritten in the form of an acceleration equation for
particles at position ${\bf{r}}$:
\begin{equation}
\label{acceleration}
\dot{{\bf{v}}}_{\alpha} = -\tilde{H}{\bf{v}}_{\alpha} - {\bf{\nabla
}}\frac{\tilde{G}_{\alpha}{m}_{\alpha}}{r} \,.
\end{equation}
The acceleration equation (\ref{acceleration}) contains all the key ingredients that affect
structure formation and it is therefore important to keep it in mind for a better intuitive understanding on the results presented in the next sections.
These ingredients are:
\begin{enumerate}
 \item a fifth force ${\bf{\nabla }} \left[\Phi_\alpha + \beta \phi
\right]$ with an effective $\tilde{G}_{\alpha} = G_{N}[1+2\beta^2]$ ;
\item a velocity dependent term $\beta \dot{\phi}{\bf{v}}_{\alpha}$ which may act as dragging or friction according to the sign of $\dot{\phi}$, replacing effectively $H \rightarrow \tilde{H} = H - \beta \dot{\phi}$;
\item a time-dependent mass for each particle $\alpha$, evolving according to
(\ref{eq:nu_mass}).
\end{enumerate} 
We use here the cosmological background mass, while for a more accurate treatment $m_{\alpha }$ would also be space dependent according to the local value of $\phi$.
The same holds for $\dot{\phi }$. A relativistic generalization of eq.(\ref{euler}) is presented in the appendix but not used in the present paper.

In a Newtonian approximation $\beta\,\delta\phi$ exceeds the gravitational potential by a factor $2\beta^2 \sim 5 \times 10^3$ in the model under investigation. A local value $\beta\,\delta\phi = 1$ means that the local mass of the neutrinos is smaller by a factor $e^{-1}$ as compared to the cosmological neutrino mass. As shown in \cite{Mota_etal_2008, Wintergerst_etal_2010, Pettorino_etal_2010} one expects substantial modifications of linear theory once $\beta\,\delta\phi$ reaches values of order unity.
Indeed non-relativistic neutrinos cluster under the effect of the fifth force, at scales estimated to be around a few $10-100$ Mpc \citep{Mota_etal_2008}. At the end, they may form stable neutrino lumps of the type described by \citet{Brouzakis_Tetradis_Wetterich_2008, Bernardini_Bertolami_2009}. 

When perturbations become non-linear, at a redshift $z \sim 1-2$, difficulties
in evaluating the characteristic features of the lumps increase. Yet, it is
crucial to estimate the size and time evolution of the gravitational potential
of the neutrino lumps. The gravitational potential influences the low angular
momenta of the CMB via the ISW \citep{Pettorino_etal_2010} and determines the
size of the cross-correlation between CMB and LSS. As a general rule of thumb,
cosmologically avaraged potentials larger than $\sim 10^{-5}$ at length scales
of 100 Mpc or more give too strong an effect on the CMB. If substantially bigger
than $10^{-5}$, the gravitational potential can also potentially erase baryonic
acustic oscillations from the CDM power spectrum, thus constraining the range of
values allowed for the coupling to match current available data
\citep{Brouzakis_etal_2011}. 
On the other hand, potentials around $10^{-5}$ or somewhat smaller may cause an observational
ISW effect, perhaps even with oscillatory structure.
Furthermore, large neutrino lumps and a rapidly evolving $\Phi_{\nu}$ may
correspond to large peculiar velocities, possibly reconciling \citep[as
suggested by ][]{Ayaita_etal_2009} the presently observed large bulk flows
\citep{Watkins_etal_2009} with observational bounds on the matter fluctuations
at similar scales \citep{Reid_etal_2010}. 
Growing Neutrino scenarios were also shown to be one of the first examples of a
realistic large backreaction mechanism which modifies the parameters of the
homogeneous and isotropic background field equations by order one effects
\citep{Pettorino_etal_2010, Nunes_etal_2011}. 

A first study of the gravitational potential for individual neutrino lumps was performed in \cite{Wintergerst_etal_2010} solving non-linear Navier-Stokes fluid equations for isolated lumps. This method allowed to follow the initial growth of non-linear fluctuations in physical space, by integrating numerically on a 3D grid nonlinear evolution equations, until virialization naturally occurs. Neutrino lumps were shown to form and mimic very large cold dark matter structures, with a typical local gravitational potential $\Phi _{\nu }\sim 10^{-5}$ for a lump size of $\sim 10$ Mpc, reaching larger values for larger lumps. 

In \cite{Pettorino_etal_2010} it was shown that a simple extrapolation of linear perturbations similar to \cite{Franca_etal_2009} to redshift $z=0$ would incorrectly overcome even the extreme bound in which all available cosmic neutrinos collapse within one single lump; this limit value reaches at most $\Phi _{\nu }\sim 10^{-3}$ for scales of about $k\sim 3 \times 10^{-3}$ Mpc$^{-1}$. In a more realistic scenario, in which neutrinos distribute within different lumps and merging is active, a  substantially lower value of $\Phi _{\nu }$ is expected. 

In the absence of accurate results for  $\Phi_\nu(k)$, a first estimate of the plausible overall size of $\Phi_\nu$ and the ISW effect was performed in \cite{Pettorino_etal_2010} by imposing reasonable bounds on the growth of fluctuations, with a careful matching between linear and non-linear analysis. As no information on the statistical abundance of neutrino lumps was available, non-linear estimates had to rely on reasonable guesses on the neutrino statistical distribution in more lumps of one or two fixed typical sizes. Interestingly, these analysis had shown some hints of oscillating patterns possibly affecting the gravitational potential. The ISW-effect on the CMB power spectrum hinted to the presence of oscillatory features, reflecting oscillations in the cosmological neutrino energy density as visible in Fig.~\ref{fig_1_const}. One may speculate that this could be responsible for small ``enhancements'' and ``depressions'' in the observed CMB spectrum at low multipoles, though the size and details of the neutrino-induced ISW-effect depend on the particular model and require a more reliable estimate of the gravitational potential.

In \cite{Wintergerst_pettorino_2010} spherical collapse was also attempted, showing the presence of a clear oscillatory pattern in the extrapolated linear density at collapse.  Non-linear power spectra of CDM and neutrino perturbations in Growing Neutrino models were also shown in \cite{Brouzakis_etal_2011}, using the approach of  \cite{Pietroni_2008}, named Time Renormalization Group, or TRG.
Such non-linear quantitative treatments cannot, however, account for a slowing down of the growth as compared to the linear approximation and only temptative reasonable bounds can be formulated: large uncertainties remain when exploring the region in which the neutrino sector becomes highly non-linear ($z \lesssim 2$).

In the next sections we present for the first time results from N-body simulations of Growing Neutrino cosmologies, following the evolution of neutrino lumps and their merging through cosmic history until $z \sim 1$. We show that the evolution of neutrino structures is initially seeded by CDM filaments, that form first as in a standard cosmological scenario. Then, the growth of neutrino lumps indeed proceeds in a clear oscillatory fashion, more pronounced than ever expected in previous works. As we will see in the following, this is also due to the fact that N-body simulations finally allow us to take into account the effects due to the directionality of velocity-dependent terms like the one in eq.(\ref{acceleration}), not included within a spherical collapse picture: switching from a `friction'-type to `dragging'-type term is determinant to enhance the oscillating evolution of neutrino lumps. Furthermore, we are able for the first time to evaluate the statistical abundance of neutrino lumps as a function of their mass.

\section{N-body simulations of Growing Neutrino cosmologies} \label{sec:nb}

\begin{table*}
\begin{tabular}{cccccccc}
Model & \begin{minipage}{50pt}
\center
Box Size \\ 
(Mpc / h)
\end{minipage} &  
\begin{minipage}{50pt}
\center
Neutrino \\ Particles
\end{minipage} &
\begin{minipage}{50pt}
\center
CDM \\ Particles
\end{minipage} &
\begin{minipage}{50pt}
\center
Neutrino part. \\ mass at $z=0$ \\ (M$_{\odot}$/h)
\end{minipage} &
\begin{minipage}{50pt}
\center
CDM part. \\ mass \\ (M$_{\odot}$/h)
\end{minipage} &
\begin{minipage}{50pt}
\center
Neutrino grav. \\
softening \\
(kpc / h)
\end{minipage} &
\begin{minipage}{50pt}
\center
CDM grav.\\
softening \\

\end{minipage} \\
\hline
$\Lambda $CDM-a& 40 & $64^3$ & $128^3$ & -- & $2.0\times 10^{9}$ & -- & $8.0$ \\
Cnu-a & 40  & $64^3$ & $128^3$ & $8.7\times 10^{9}$ & $2.0\times 10^{9}$ & $15.0$ & $8.0$ \\
$\Lambda $CDM-b& 120 & $64^3$ & $128^3$ & -- & $5.4\times 10^{10}$ & -- & $24.0$ \\
Cnu-b & 120  & $64^3$ & $128^3$ & $2.3\times 10^{11}$ & $5.4\times 10^{10}$ & $45.0$ & $24.0$ \\
$\Lambda $CDM-c& 320 & $128^3$ & $256^3$ & -- & $1.3\times 10^{11}$ & -- & $16.0$ \\
Cnu-c & 320  & $128^3$ & $256^3$ & $5.6\times 10^{11}$ & $1.3\times 10^{11}$ & $30.0$ & $16.0$ \\
\hline
\end{tabular}
\caption{The different simulations considered in the present work for Growing Neutrino cosmologies and for the standard $\Lambda $CDM model.}
\label{tab:simulations}
\end{table*}
In order to study the nonlinear evolution of structures in the context of Growing Neutrino cosmologies, 
we run a series of N-body simulations that take into account the interaction of neutrinos with the cosmon
and all the related effects on the growth of density perturbations that have been discussed in the previous Sections.
To this end, we have made use of the modified version by \citet{Baldi_etal_2010} of the cosmological N-body code {\small GADGET-2} \citep{gadget-2}, 
that was specifically designed to include the effects of interacting DE models.

The only modification that has to be implemented in the code with respect to the algorithm described in \citet{Baldi_etal_2010}
concerns the transition of the neutrino particles from the relativistic to the non-relativistic regime. Due to the fast growth of the neutrino mass, 
this transition can be roughly approximated with a step function for the neutrino equation of state, sharply changing from $w_{\nu} = 1/3$ to $w_{\nu }=0$
at the transition redshift $z_{{\rm nr}}$. In this study we have assumed $z_{{\rm nr}} = 4$ based on the results of the background 
evolution of the specific Growing Neutrino model under investigation described in Sec.~\ref{sec:gnc}. 
According to this approximation, neutrino particles in the simulation box are free-streaming at all scales for $z>z_{{\rm nr}}$, when their mass is negligible with respect to the CDM particle mass, while they behave as cold dark matter for $z<z_{\rm nr}$ when 
their growing mass rapidly makes them 
non-relativistic. To implement the transition in our numerical machinery, we have artificially switched off the gravitational
interaction for neutrino particles before $z_{{\rm nr}}$, and set their mass to zero. Therefore, the neutrino particles in the simulation box will neither experience any acceleration nor exert any force on the CDM particles during their relativistic regime $z>z_{{\rm nr}}$, and will therefore move on straight trajectories with their own initial velocities. 
The initial velocities are set in the simulations initial conditions by randomly sampling a Fermi-Dirac distribution with a temperature corresponding to the average neutrino temperature at $z_{{\rm nr}}$ according to our model of growing neutrino mass.
This implementation ensures that the neutrino density field will remain completely homogeneous in the simulation box until $z_{{\rm nr}}$, and that the velocity 
distribution of the neutrino particles at $z_{{\rm nr}}$ corresponds to the correct Fermi-Dirac velocity distribution for a neutrino with mass $m_{\nu }(z_{{\rm nr}})\approx 0.02$ eV.

After the transition redshift $z_{{\rm nr}}$, neutrino masses are set to their correct value, according to the neutrino mass evolution of the specific model
under investigation, and the gravitational interaction as well as the scalar fifth-force are switched on. 
Therefore, right after $z_{{\rm nr}}$ neutrino particles
will start to experience gravitational accelerations sourced by the surrounding CDM density field, that by $z_{{\rm nr}}$ will have already developed significant inhomogeneities. 
Furthermore, neutrino particles
will also start exerting gravitational forces on other CDM and neutrino particles, due to their rapidly growing mass.
It is however the fifth-force acting among neutrino particles that soon dominates over gravity, driving the large scale evolution of neutrino structures at $z < z_{{\rm nr}}$.
\ \\

With this implementation we have run a series of N-body simulations of structure formation for the specific Growing Neutrino model described in Sec.~\ref{sec:gnc}.
These consist in cosmological boxes of different sizes containing both neutrino and CDM particles. The initial conditions for the CDM 
component have been generated by setting up a random-phase realization of the Eisenstein \& Hu power spectrum \citep{Eisenstein_Hu_1997} according to the
Zel'dovich approximation \citep{Zeldovich_1970}. The normalization amplitude of the power spectrum is adjusted to a value of $\sigma _{8} = 0.769$.
The displacements of the particles are then rescaled to the starting redshift of the simulation $z_{i} = 60$ with the linear growth factor $D_{+}$ which is assumed to be
the $\Lambda $CDM one. Due to the impact of the growing neutrino component on the growth of structures, however, the real CDM growth factor might slightly differ from the 
standard $\Lambda $CDM one, thereby driving to an actual value of $\sigma _{8}$ different from the one used for the normalization of the initial conditions.
Since the neutrino component 
is assumed to be completely homogeneous before $z_{{\rm nr}}$, no displacement is applied to neutrino particles.
The procedure adopted to generate the initial velocities for the neutrino component has already been described above. 

\begin{figure}
\includegraphics[scale=0.4]{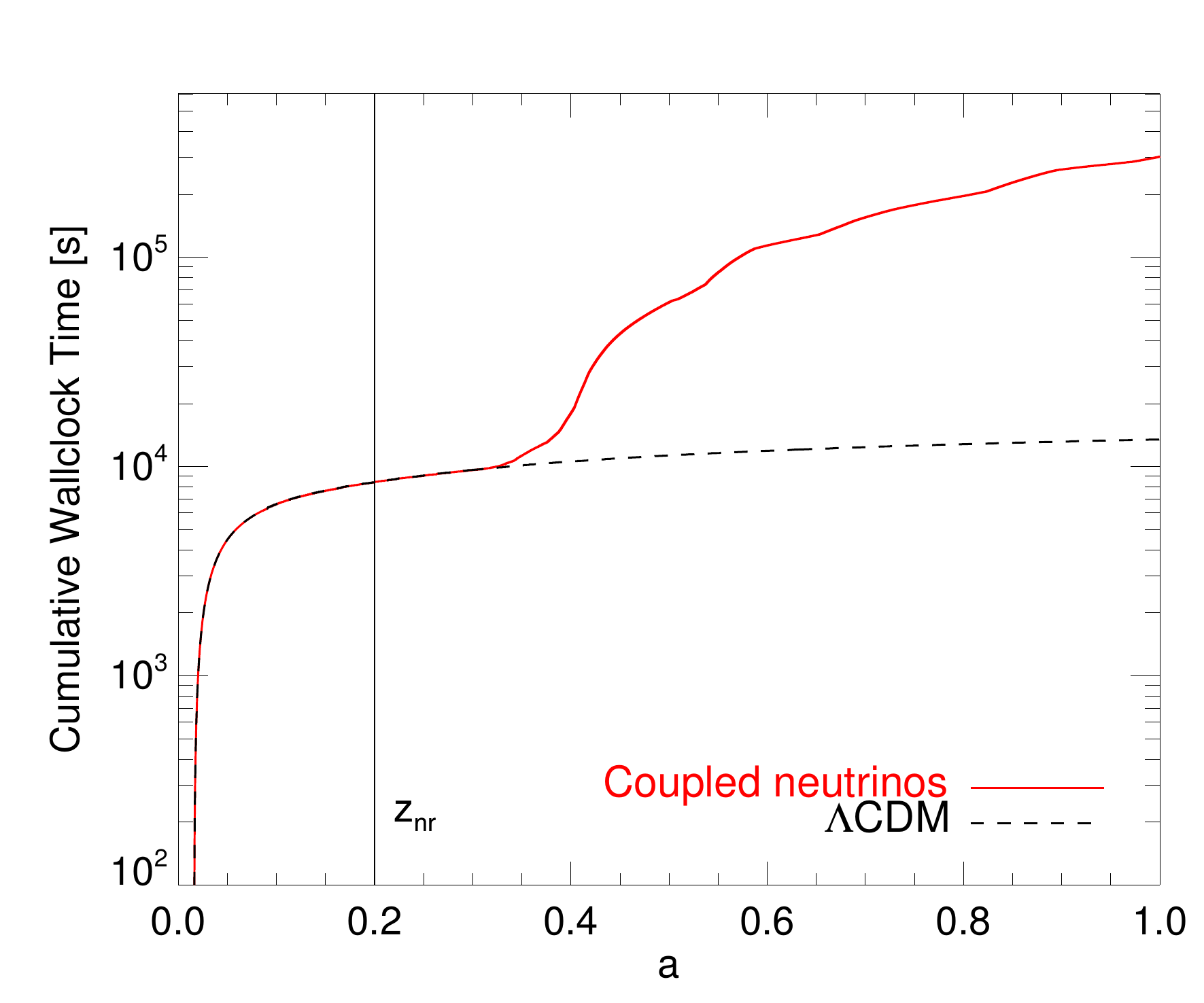}
\caption{The cumulative wallclock time (in seconds) as a function of the cosmological scale factor $a$ for a standard $\Lambda $CDM simulation (black dashed line) and a Growing Neutrino simulation with the same box size and particle number. The vertical line indicates the transition redshift $z_{\rm nr}$ at which the coupling between DE and massive neutrinos becomes active.}
\label{fig:cpu_time}
\end{figure}

A major numerical challenge for the study of Growing Neutrino cosmologies by means of N-body simulations consists in the significant 
increase of the overall computational time compared to standard CDM simulations with the same number of particles, which can be as large as
a factor of $10^2$. This large increase of the required computational time is mainly due to the timestepping criterion, which
relates the individual timestep of a given particle to the acceleration experienced by the particle in the previous timestep, according to the equation:
\begin{equation}
\label{timestep}
\Delta t_{i} \propto \sqrt{\epsilon/|\vec{a}_{{\rm old}}|}\,,
\end{equation}
where $\epsilon $ is the gravitational softening and $\vec{a}_{{\rm old}}$ is the gravitational acceleration of the particle.
With the above timestepping criterion, as soon as the neutrino fifth-force becomes active right after $z_{\rm nr}$, the acceleration experienced by neutrino particles
exceeds by a factor $2\beta ^{2}$ the standard gravitational acceleration, and the corresponding timestep $\Delta t$ is therefore reduced by a factor $1/\sqrt{2\beta ^{2}}$,
that for the model under investigation is of the order of $\sim 80$. This in turn determines a similar increase in the total computational time of the simulations. The situation is well illustrated in Fig.~\ref{fig:cpu_time}, where the cumulative wallclock time (in seconds) for two of our simulations is plotted as a function of the cosmological scale factor $a$. The solid red line represents the simulation for coupled neutrinos, while the black dashed line is for the standard $\Lambda $CDM cosmology. 
As one can clearly see, while before the 
transition redshift $z_{\rm nr}$ the two simulations evolve with identical wallclock times, after the neutrino fifth-force becomes active at $z=z_{\rm nr}$ the coupled neutrino simulation starts deviating from the $\Lambda $CDM one, with a rapidly increasing computational cost. 
In order to alleviate this problem, since the increase of the total computational time
for coupled neutrino simulations arises from the reduced timestep of neutrino particles,  and since neutrino structures are expected to form only at relatively large scales \citep[as predicted by \eg][]{Mota_etal_2008} thereby requiring a relatively low spatial resolution, we have limited the number of neutrino particles in each simulation to be $8$ times smaller than the corresponding number of CDM particles. This contributes to reduce the increase of the total computational cost of the simulations, resulting in a relative overhead of a factor of  $\approx 25$ for our coupled neutrino runs as compared to $\Lambda $CDM (see again Fig.~\ref{fig:cpu_time}), significantly smaller than the overhead of a factor  of $80$
estimated for an equal number of CDM and neutrino particles.
\ \\

The details of the simulations considered in the present work are described in Table~\ref{tab:simulations}.

\section{Results} \label{sec:results}

We now describe the results obtained from the analysis of the different simulations described in Table~\ref{tab:simulations}.
First of all, we focus our attention on the range of validity of the N-body approach for Growing Neutrino cosmologies,
then we study the evolution and the statistical properties of neutrino structures at different scales, and finally 
we investigate the backreaction effects of the growth of neutrino structures on the CDM distribution.

\subsection{Evolution of neutrino velocities} \label{subsec:velocities}
\label{neutrino_vel}

\begin{figure*}
\includegraphics[scale=0.4]{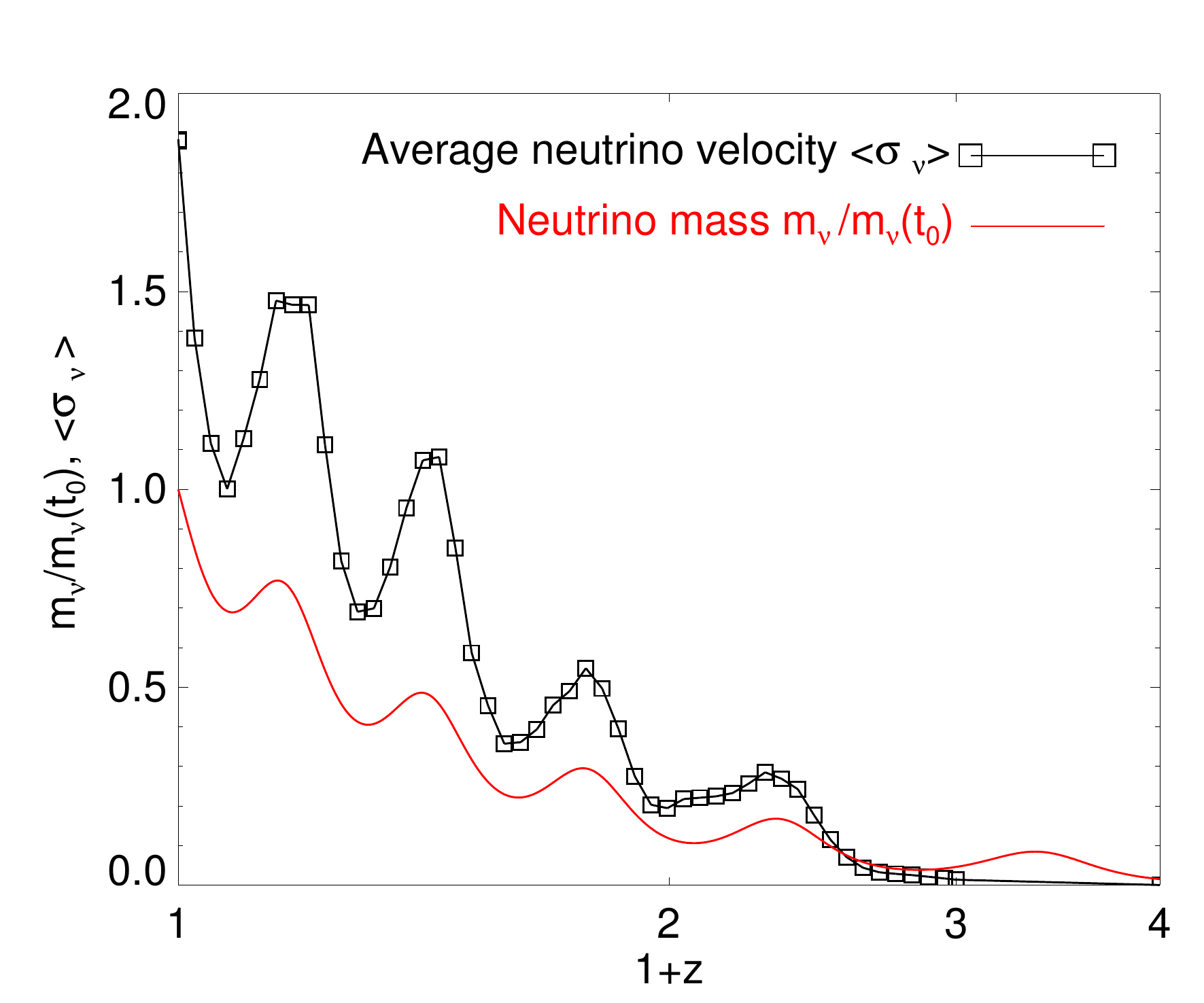}
\includegraphics[scale=0.4]{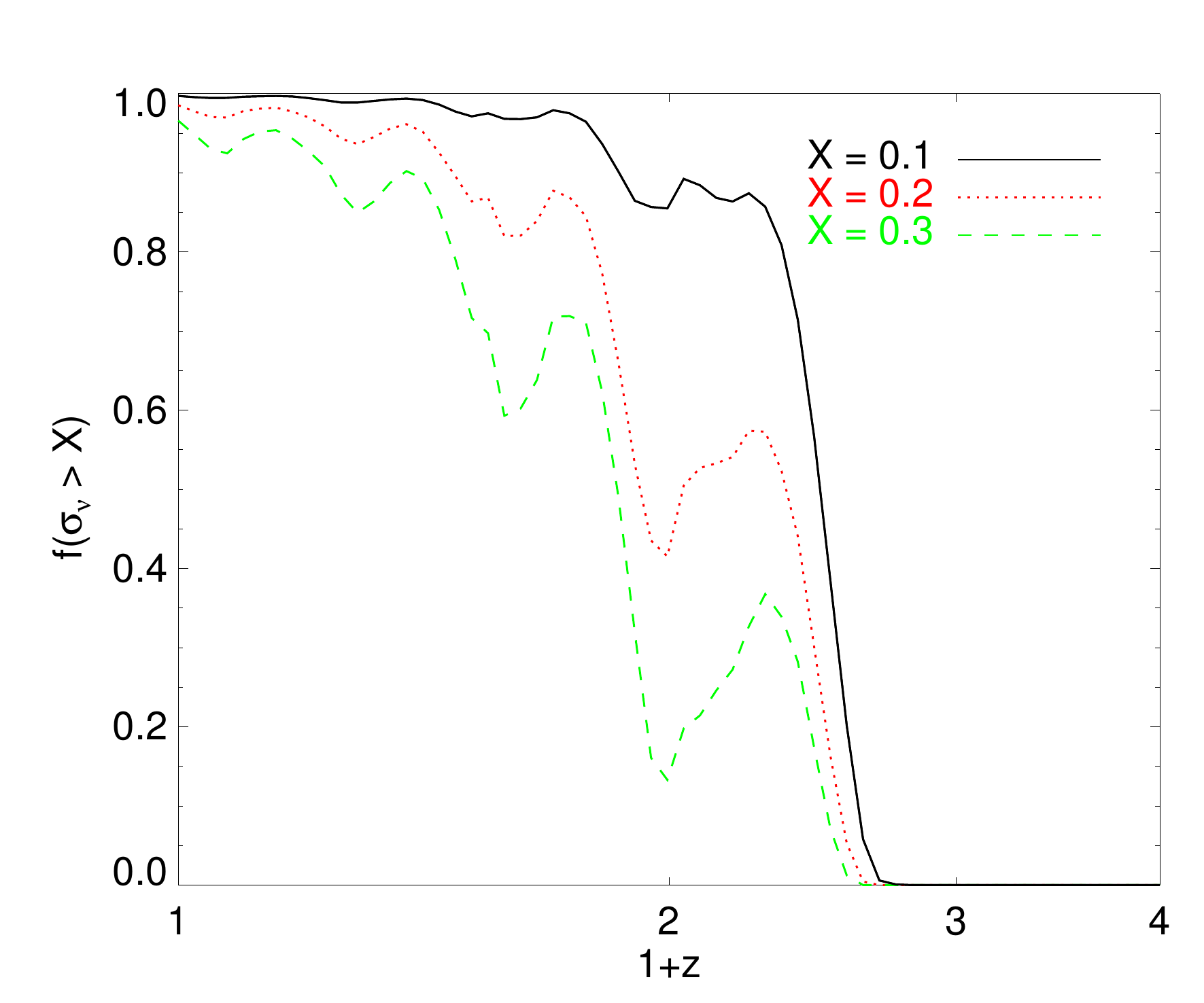}
\caption{{\em Left panel}: The evolution of the neutrino mass $m_{\nu }/m_{\nu }(t_{0})$ (solid red line) and of the average neutrino peculiar velocity in units of the speed of light $<\sigma _{\nu }>$ (black line, squares) as extracted from the Cnu-b simulation described in Table~\ref{tab:simulations}, as a function of redshift. {\em Right panel}: the fraction of neutrino particles in the Cnu-b simulation of Table~\ref{tab:simulations} exceeding a fraction $X$ of the speed of light as a function of redshift. The different curves are for $X=0.1$ (solid black), $X=0.2$ (dotted red) and $X=0.3$ (dashed green).}
\label{fig:velocities}
\end{figure*}

As a first step in our analysis we want to test the validity of the N-body approach to study the nonlinear evolution of structure formation
in the context of Growing Neutrino cosmologies.
N-body codes are designed to follow the evolution of density perturbations under gravitational instability processes by solving
the Poisson equation in the newtonian limit. This means that the scales of interest should be small compared to the cosmological
horizon and that particle velocities should be small compared to the speed of light. A third issue concerns the field dependence of the neutrino masses in our model. The cosmon field inside neutrino lumps may differ from its cosmological value, and as a consequence the neutrino mass may depend on position. This ``backreaction'' effect could be substantial for very large neutrino lumps \citep{Pettorino_etal_2010,Nunes_etal_2011}. Its computation would require to follow the scalar field distribution during the simulation, which goes beyond the scope of the present implementations of N-body codes. (In our approach we use the cosmological background value of the cosmon for the determination of the neutrino mass, such that $m_{\nu}$ depends on time but not on position).

For the first issue, \citet{Chisari_Zaldarriaga_2011}
have recently shown that even for scales comparable or larger than the cosmological horizon the standard N-body algorithms provide
a correct computation of the gravitational potential. With our simulations we can now address the second issue related to 
particle velocities. Particle velocities constitute a potential problem for N-body codes since 
in the newtonian limit particles are allowed to accelerate to arbitrarily large values, possibly even exceeding the speed of light.
While this is not a problem for standard cosmological simulations, where peculiar velocities
are always small with respect to the speed of light due to the smallness of the newtonian gravitational constant $G_{{\rm N}}$ , it might turn out to be a problem
for non-standard scenarios -- as \eg the Growing Neutrino model under investigation in this work -- where the attraction is governed by an effective strength $G_{{\rm eff}}$
that in some situations 
significantly exceeds the value of $G_{{\rm N}}$.

For the specific case of the model considered in the present study, while the gravitational dynamics of CDM particles is always
governed by the newtonian gravitational constant $G_{{\rm N}}$, the dynamical evolution of neutrino particles after the transition redshift 
$z_{{\rm nr}} = 4$ is driven by an effective coupling $G_{{\rm eff}} = G_{{\rm N}}(1 + 2\beta ^{2})$, which is more than $5\times 10^{3}$
times larger than $G_{{\rm N}}$.
It is therefore natural to ask whether this large effective gravitational constant leads to an acceleration of neutrino particles
to exceedingly high velocities, and to which extent an N-body treatment of this system can be considered correct.
In order to investigate this problem, we compute the evolution of the average neutrino peculiar velocities by taking the mean
of the neutrino particles velocities in our simulation boxes at different timesteps. The result of this procedure for one of our
simulations (the Cnu-b run of Table~\ref{tab:simulations}) is shown in the left panel of Fig.~\ref{fig:velocities}, where the solid black line and the black squares represent the
average neutrino peculiar velocity extracted from the snapshots of the simulation in units of the speed of light $<\sigma _{\nu }>$,
where $\sigma _{\nu }\equiv v_{\nu }/c$.
As expected, we find that neutrino velocities rapidly grow after the transition redshift $z_{{\rm nr}}$ reaching already at $z\sim 1$
a non-negligible fraction ($20-30\%$) of the speed of light, and exceed the value of $c$ at $z\sim 0.5$. The final value of the 
average neutrino velocity in the simulation at $z=0$ is almost two times larger than the speed of light,
having reached clearly a regime where N-body simulations are no longer reliable.

A comparison with the non-linear analysis done in \cite{Wintergerst_etal_2010} solicited by the present study has confirmed
the same trend of neutrino velocities shown in the left panel of Fig.~\ref{fig:velocities} (N.~Wintergerst, {\em private communication}).

It is therefore clear that the N-body approach is not sufficient to follow correctly the dynamical evolution of neutrino particles up to
the present time, and that relativistic corrections would be required in order to implement in the treatment the concept of a 
speed limit. This however would go beyond the newtonian limit in which N-body codes are implemented, and would require
the development of completely new algorithms. In the present work, we do not want to undertake such a challenging enterprise,
but we will rather try to assess the range of validity of the standard N-body approach and will restrict our analysis to this limited range.
In particular, as a result of our analysis shown in Fig.~\ref{fig:velocities}, we can assume our simulations to be no longer reliable 
at redshifts below $z\sim 1$. We will therefore discard all the simulations outputs at $z<1$ and will limit our analysis to the early stages of the nonlinear
evolution of neutrino large scale structures.
\ \\

The other interesting feature emerging from the analysis shown in the left panel of Fig.~\ref{fig:velocities}, is that neutrino velocities
do not grow monotonically in time, but rather exhibit an oscillatory behavior corresponding to a series of subsequent accelerations
and decelerations of neutrino particles. This peculiar evolution is not a numerical artifact, but is directly related to the evolution
of neutrino masses. As discussed in Sec.~\ref{sec:gnc}, when neutrinos become non-relativistic the coupling with the cosmon gets active
and the effective potential $V_{{\rm eff}}$ for the scalar field acquires a minimum where the scalar field stops, thereby mimicking a cosmological constant.
However, due to its initial kinetic energy, before stopping the scalar field oscillates around the minimum of $V_{{\rm eff}}$, and these
oscillations determine a corresponding oscillation of the neutrino mass according to Eq.~\ref{eq:nu_mass}.

In the left panel of Fig.~\ref{fig:velocities}, the solid red line represents the evolution of the neutrino mass $m_{\nu }/m_{\nu }(t_{0}) \sim \exp(-\beta \phi /M_{{\rm Pl}})$.
As it clearly appears from the plot, the oscillations of the average neutrino velocity $<\sigma _{\nu }>$ are correlated with the oscillations
in the neutrino mass. This result can be easily interpreted by having a look at the neutrino acceleration equation, Eq.~\ref{acceleration}.

Furthermore, we may also infer from Fig.~\ref{fig:velocities} that even for a large present average neutrino mass $m_{\nu }(t_{0}) = 2.1$ eV, the neutrino mass at $z=1$ is $m_{\nu }\sim 0.3 - 0.4$ eV,
with even smaller values for $z>1$. It is yet to be investigated whether this value of $m_{\nu}(t_{0})$ remains compatible with restrictions from structure formation \citep[][]{wmap7}.  

The right panel of Fig.~\ref{fig:velocities} shows -- for the same simulation Cnu-b -- the fraction of neutrino particles whose velocity exceeds a given fraction $X$ of the speed of light.
As one can see from the plot, almost $90\%$ of particles (solid black line) have velocities larger than $10\%$ of the speed of light at $z\sim 1$, while $20-35\%$ of
particles (dashed green line) exceed $30\%$ of the speed of light in the same redshift range.

\subsection{Large scale neutrino structures} \label{subsec:oscillations}
\label{lss}

\begin{figure*}
\includegraphics[scale=0.3]{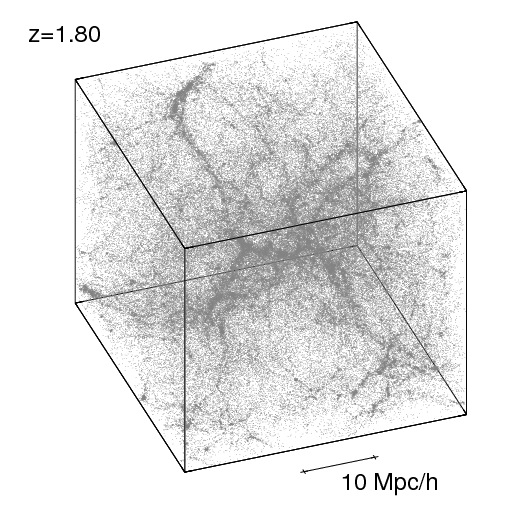}
\includegraphics[scale=0.3]{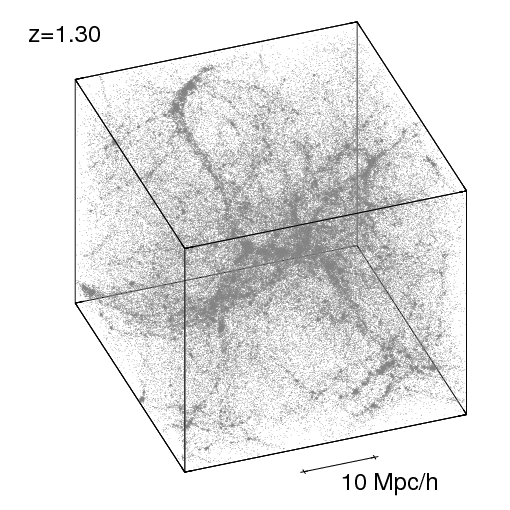}
\includegraphics[scale=0.3]{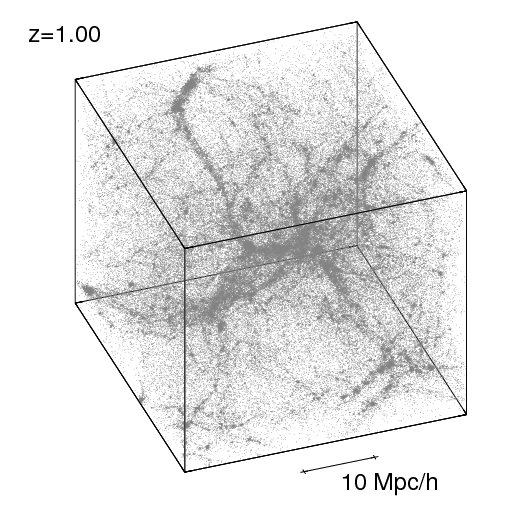}\\
\includegraphics[scale=0.3]{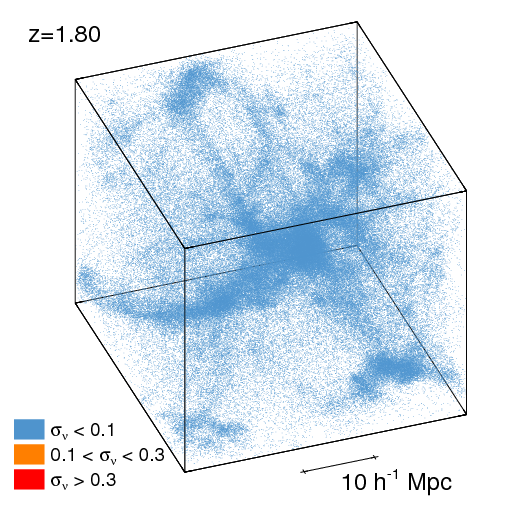}
\includegraphics[scale=0.3]{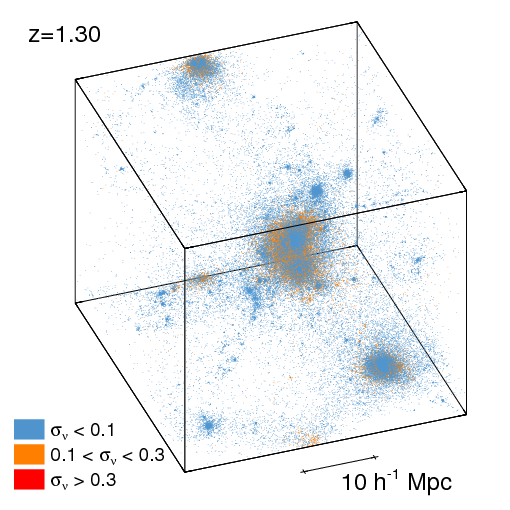}
\includegraphics[scale=0.3]{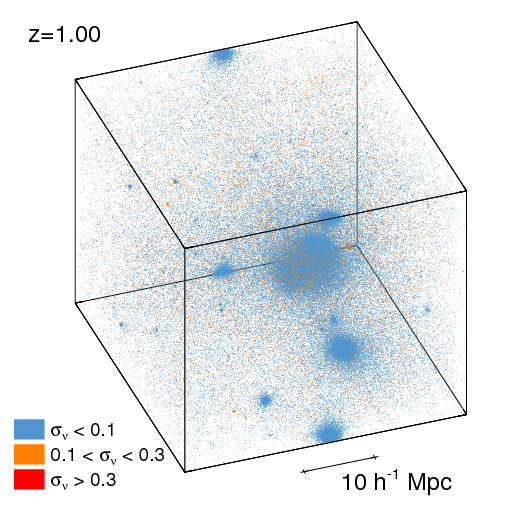}\\
\caption{The distribution of CDM (top row) and neutrino (bottom row) particles in the Cnu-a simulation at $z=1.8$ (left) $z=1.3$ (middle) and $z=1$ (right). The neutrino particles are colored according to their velocity $\sigma _{\nu}$, as explained in the legend.}
\label{fig:boxes1}
\end{figure*}
\normalsize

\begin{figure*}
\includegraphics[scale=0.3]{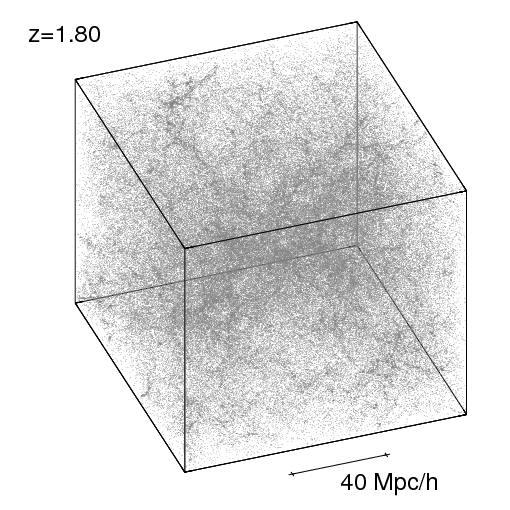}
\includegraphics[scale=0.3]{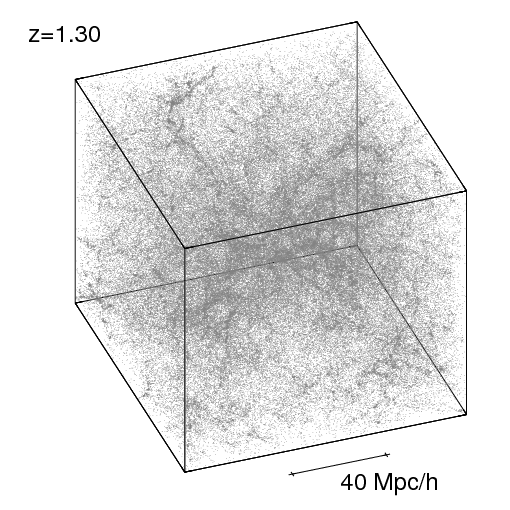}
\includegraphics[scale=0.3]{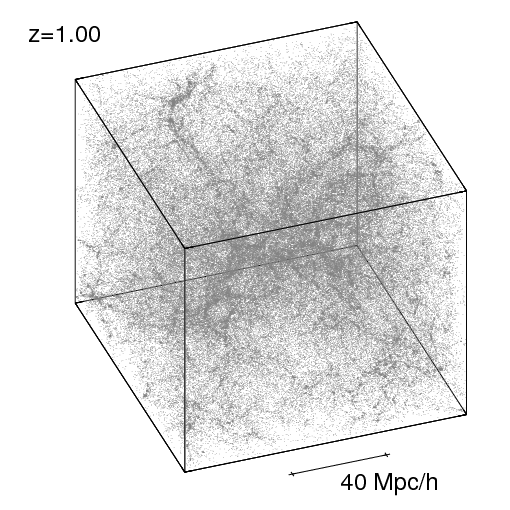}\\
\includegraphics[scale=0.3]{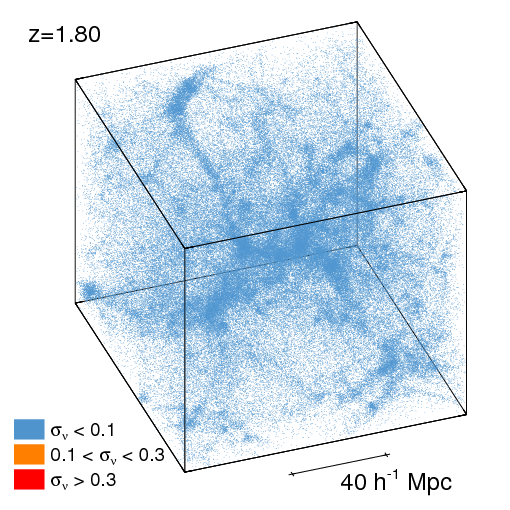}
\includegraphics[scale=0.3]{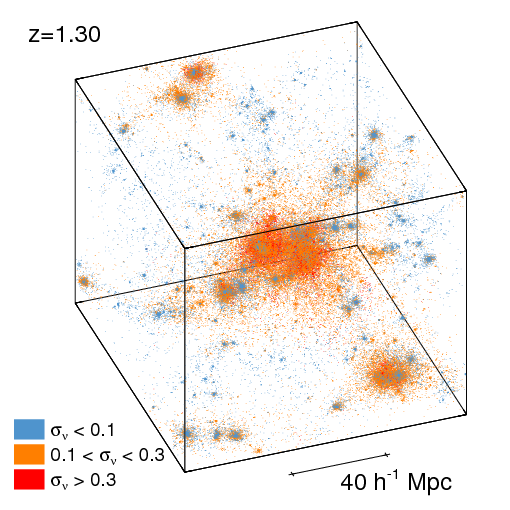}
\includegraphics[scale=0.3]{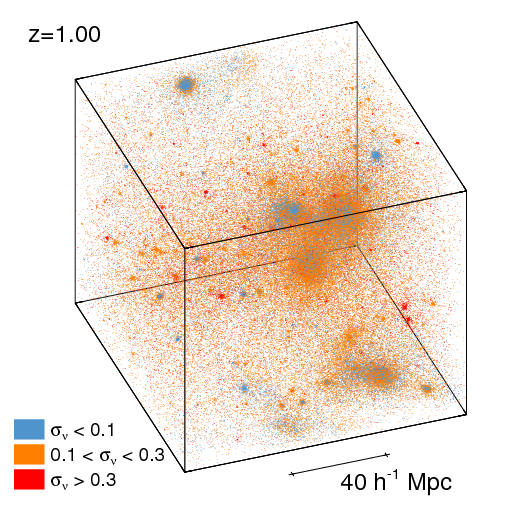}
\caption{The same as for Fig.~\ref{fig:boxes1} but for the Cnu-b simulation of Table~\ref{tab:simulations}.}
\label{fig:boxes2}
\end{figure*}
\normalsize

\begin{figure*}
\includegraphics[scale=0.3]{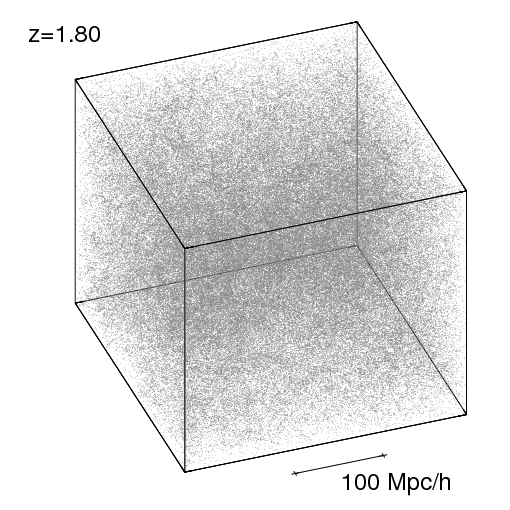}
\includegraphics[scale=0.3]{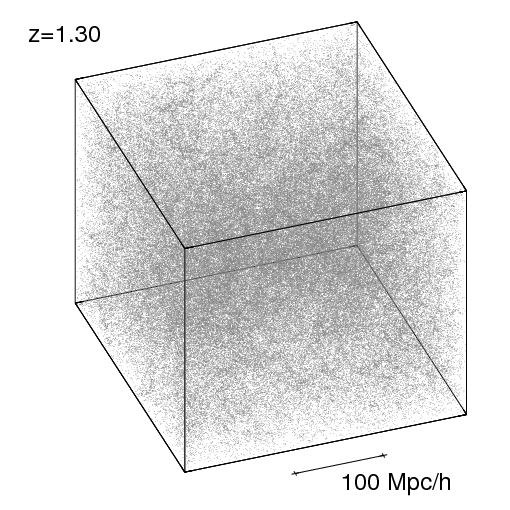}
\includegraphics[scale=0.3]{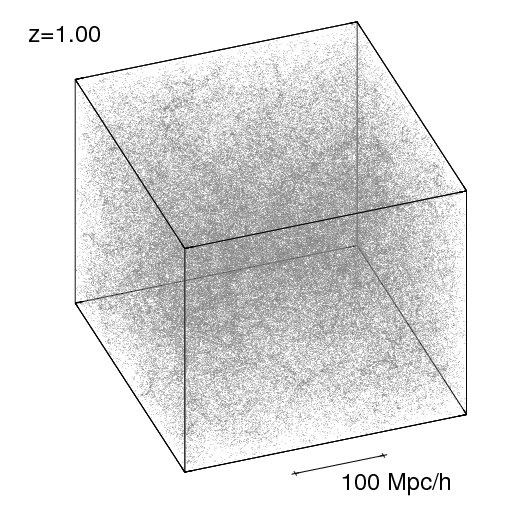}\\
\includegraphics[scale=0.3]{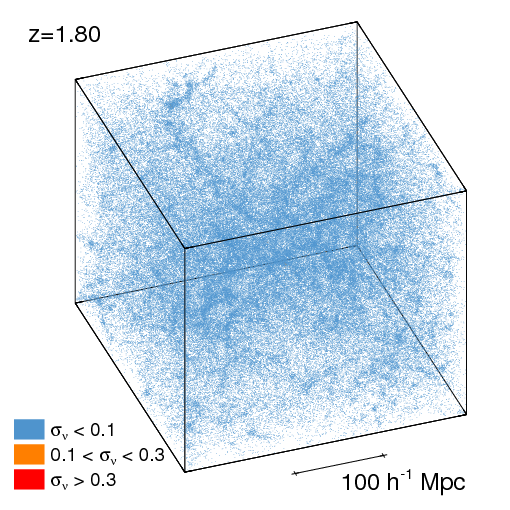}
\includegraphics[scale=0.3]{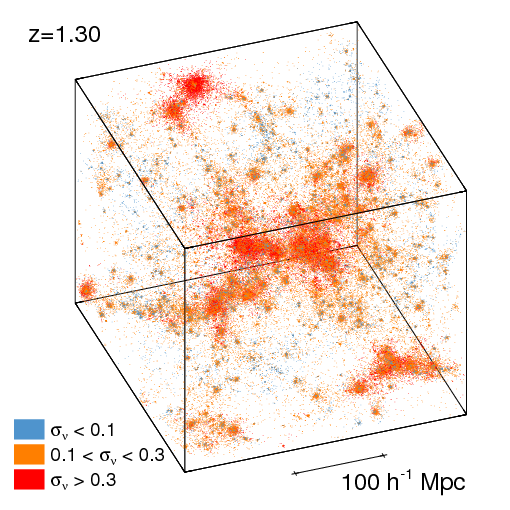}
\includegraphics[scale=0.3]{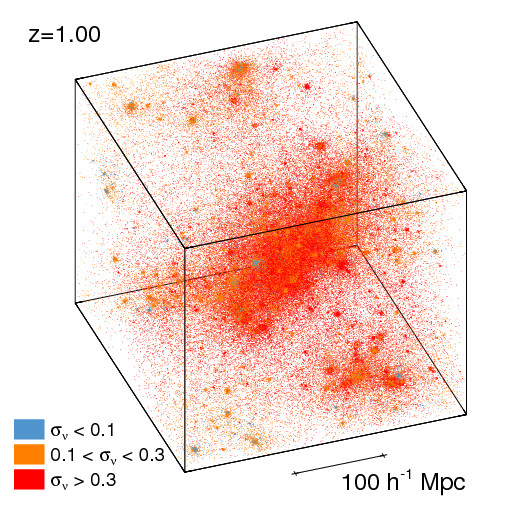}\\
\caption{The same as for Fig.~\ref{fig:boxes1} but for the Cnu-c simulation of Table~\ref{tab:simulations}.}
\label{fig:boxes3}
\end{figure*}
\normalsize

Having in mind the limitations of the N-body approach for the study of coupled neutrino cosmologies discussed in the previous Section,
we now start to investigate the growth of neutrino and CDM structures in our simulations. 
We begin by visualizing the evolution of the neutrino and CDM particle distributions in all the simulations described in Table~\ref{tab:simulations}
for three different redshifts above our limiting redshift $z=1$ and below the transition redshift $z_{{\rm nr}} = 4$.
Fig.~\ref{fig:boxes1}, Fig.~\ref{fig:boxes2}, Fig.~\ref{fig:boxes3} show a random subsample of CDM (grey points, top rows) and neutrino (colored points, bottom rows)
particles at redshifts $z=1.8$ (left columns), $z=1.3$ (central columns), and $z=1.0$ (right columns) for our coupled neutrino simulations
Cnu-a (Fig.~\ref{fig:boxes1}), Cnu-b (Fig.~\ref{fig:boxes2}) and Cnu-c (Fig.~\ref{fig:boxes3}).
In these figures, neutrino particles have been colored according to their velocities, with blue points representing ``slow" particles ($\sigma _{\nu } < 0.1$),
orange points representing ``intermediate velocity" particles ($0.1 < \sigma _{\nu } < 0.3$) and red points representing ``fast" particles ($\sigma _{\nu} > 0.3$).
\ \\

By having a look at the Cnu-a simulation (Fig.~\ref{fig:boxes1}) one can clearly see how CDM structures (first row)
evolve on scales of the order of few tens of Mpc. The main filamentary structure is already present at $z=1.8$ and evolves
by growing overdensities and enlarging voids up to $z=1$. It is very interesting to see what happens to the neutrino distribution (second row)
at the same scales and in the same redshift range. At $z=1.8$ neutrinos have a mass of $\approx 0.2$ eV and had
sufficient time after the transition redshift $z_{{\rm nr}}$ to fall into the potential wells of the already developed CDM structures. In fact,
as one can clearly see in the figure, the distribution of neutrino particles at $z=1.8$ roughly follows the underlying CDM distribution, with the 
main concentrations of neutrinos located in the same positions as the underlying CDM halos and filaments.
At this epoch, however, neutrinos did not have enough time to accelerate to very high velocities, and in fact the figure shows that basically
none of the neutrino particles has a velocity $\sigma _{\nu }$ larger than $0.1$. 

The situation is already significantly different at $z=1.3$, where
the first isolated neutrino lumps have formed, and where the distribution of neutrino particles does not trace anymore the underlying 
CDM filamentary structure, but rather concentrates in a few very large and isolated halos. At this time, some particles with higher velocities 
are clearly visible in the central regions of the largest neutrino lumps, while there is almost no diffuse neutrino component left over outside the main bound structures:
practically all neutrinos are residing in bound halos at this redshift.

The situation at $z=1$ therefore appears 
at a first glance quite surprising: while the number of large neutrino halos in the box has been reduced to a few by the merging 
of a large number of smaller neutrino lumps, a significant diffuse component of free neutrinos, that was 
completely absent at $z=1.3$, surprisingly appears again. We will find this phenomenon also in the larger scale simulations Cnu-b and Cnu-c,
and we will come back later on its interpretation. The neutrino distribution at $z=1$ in this small scale simulation shows the presence
of a few neutrino lumps with a scale of $5-10$ Mpc/h in a volume of $6.4\times 10^{4} ($Mpc/h$)^{3}$, and a dominant fraction of ``slow" neutrinos,
in agreement with Fig.~{\ref{fig:velocities}}.
\ \\

If we now consider the Cnu-b simulation (Fig.~\ref{fig:boxes2}) we can see how the CDM structures 
are significantly less evolved (as expected) on these larger scales as compared to the Cnu-a simulation. However, the large scale filaments and the main
CDM structures are visible already at $z=1.8$, and become progressively better defined as time goes by until $z=1$.
Also in this case, however, the neutrino distribution at $z=1.8$ seems to follow the main features of the CDM structures, with
neutrino particles distributed in very large scale lumps and filaments, and with still low velocities as basically all the neutrinos appear as ``slow" particles
at this stage. 

At $z=1.3$ the situation is then completely different, with a large number of neutrino structures at very different scales, and with
the presence of ``slow" neutrinos as well as ``intermediate" and ``fast" neutrinos.
As already found for the Cnu-a simulation, also in this case most of the neutrinos are residing in lumps, and there is basically
no diffuse component of free neutrinos at this stage, \ie the space between neutrino lumps is practically empty of neutrinos. 

Once again, the situation changes at $z=1$, where the number of neutrino structures has 
significantly reduced, and one can identify neutrino halos on scales ranging from a few Mpc/h up to $20-30$ Mpc/h. Also in this case, 
besides a few lumps that are still surrounded by ``slow" neutrinos and a majority of lumps made by ``intermediate" and ``fast" neutrinos,
a significant diffuse component of free neutrinos appears again, after being 
absent at $z=1.3$.
\ \\

We can now move to consider the largest scale simulation at our disposal, the simulation Cnu-c (Fig.~\ref{fig:boxes3}).
At these very large scales ($320$ Mpc/h aside), the CDM distribution (upper row) appears  very homogeneous (as expected)
and no significant structures can be easily identified even at $z=0$ at our resolution. Nevertheless, the CDM gravitational potential is 
sufficiently evolved, even at these large scales, to source the growth of neutrino overdensities right after the transition redshift $z_{{\rm nr}}$
such that at $z=1.8$ the neutrino distribution already shows a significant level of structure that is not visible in the underlying CDM distribution.
This shows very clearly how the neutrino distribution evolves in the context of Growing Neutrino scenarios:
at $z_{{\rm nr}}$, the neutrino component is completely homogeneous and suddenly starts to experience accelerations sourced
by the gravitational potential of the underlying CDM distribution; as soon as neutrinos, as a consequence of these accelerations,
start to move towards the minima of the CDM gravitational potential and develop their own inhomogeneities, the fifth-force
acting between neutrino particles and mediated by the scalar field $\phi $, that is $5\times 10^{3}$ times stronger than gravity, starts
driving the evolution of neutrino overdensities that can therefore grow much faster than the underlying CDM density field, 
and develop significant structures at scales where CDM appears to be almost completely homogeneous. 
The neutrino distribution at $z=1.8$ in Fig.~\ref{fig:boxes3} therefore shows very clearly how the CDM density field acts as a ``seed" for the 
growth of neutrino lumps. 

At this stage, also on these large scales, the neutrino component is almost completely made of ``slow" particles, while 
at $z=1.3$ a large number of isolated neutrino lumps on a wide range of scales have already formed and are mainly composed
of ``intermediate" and ``fast" neutrinos. Once more, there is basically no diffuse component of free neutrinos at $z=1.3$ and
all neutrinos are residing in lumps, but the situation changes at $z=1$, where a large number of neutrino lumps is still visible but
where a significant diffuse component of free neutrinos appears again. In this case, the large majority of neutrino particles at $z=1$
is ``fast" while very few ``slow" neutrinos are visible in the simulation box.\\

The analysis of the evolution of the neutrino distribution in these three simulations of different scales that was described above, represented in
Fig.~\ref{fig:boxes1}, Fig.~\ref{fig:boxes2} and Fig.~\ref{fig:boxes3}, clearly identifies two interesting and distinct phenomena.

First, if we look at the neutrino distribution at $z=1$ in the three simulations, we can notice a clear difference: the neutrino
component is mainly composed of ``slow" neutrinos in the Cnu-a simulation ($40$ Mpc/h aside), of ``intermediate" neutrinos
in the Cnu-b simulation ($120$ Mpc/h aside), and of ``fast" neutrinos in the Cnu-c ($320$ Mpc/h aside). This indicates that the inclusion
of larger scale modes in the CDM density field determines a larger integrated acceleration for the neutrino particles, since neutrinos 
have more time to accelerate in the large scale linear CDM potential than in the small scale nonlinear one. 

Besides possible questions of numerical convergence, the sequence of different
scales for the boxes indicates an issue for the scaling of the simulations.
While we can find the structures of the small scale simulations also in the
simulations at larger box size, the typical velocities for structures of a given
size are not the same. In Fig.\ref{fig:boxes1} the largest structures have
typical size of a few Mpc/h, with nonrelativistic velocities of the
``particles''. Structures of a similar size in
Fig.\ref{fig:boxes2},\ref{fig:boxes3} have typically higher velocities. The
reason seems simple: the lumps of say $5$ Mpc/h size move as whole with a high
peculiar velocity in the simulations with large box size, while they feel no
``outside attraction'' in the box of a comparable size and do therefore not
develop a collective velocity. In the three simulations no saturation of this
phenomenon is seen at the largest box size. Investigating still larger box sizes
is presumably not of much help since the peculiar velocities become so large
that a nonrelativistic approximation seems no longer meaningful. It is not known
how the collective velocities and scattering of smaller size structures
influence precisely the neutrino distributions inside these structures.The
conclusion drawn from our simulations should therefore only be considered as
semi-quantitative. Nevertheless, a few interesting lessons can be drawn.

\subsection{Oscillating structure formation}

We have found in all the three simulations, independently on the size of the simulation box, that the neutrino distribution
has a significant diffuse component at $z=1.8$, then shows basically no diffuse component at $z=1.3$, when all neutrinos are residing in lumps,
and finally a significant diffuse component appears again at $z=1$. This means that a large number of neutrino particles that were residing in
gravitationally bound lumps at $z=1.3$ are ``released" and they appear no longer bound at $z=1$. This effect seems surprising since 
in standard gravitational instability processes for collisionless dynamics, bound objects remain bound and the diffuse component of particles
continuously decreases. In other words, in the standard picture of gravitational instability density fluctuations can only grow in time.
In this case, instead, we are witnessing an oscillatory behavior of the gravitational instability of neutrino overdensities that first collapse
into a collection of isolated bound objects, separated by large empty regions, and then decay again releasing a significant fraction of their
content into a new homogeneous diffuse neutrino field. 

\begin{figure}
\includegraphics[scale=0.45]{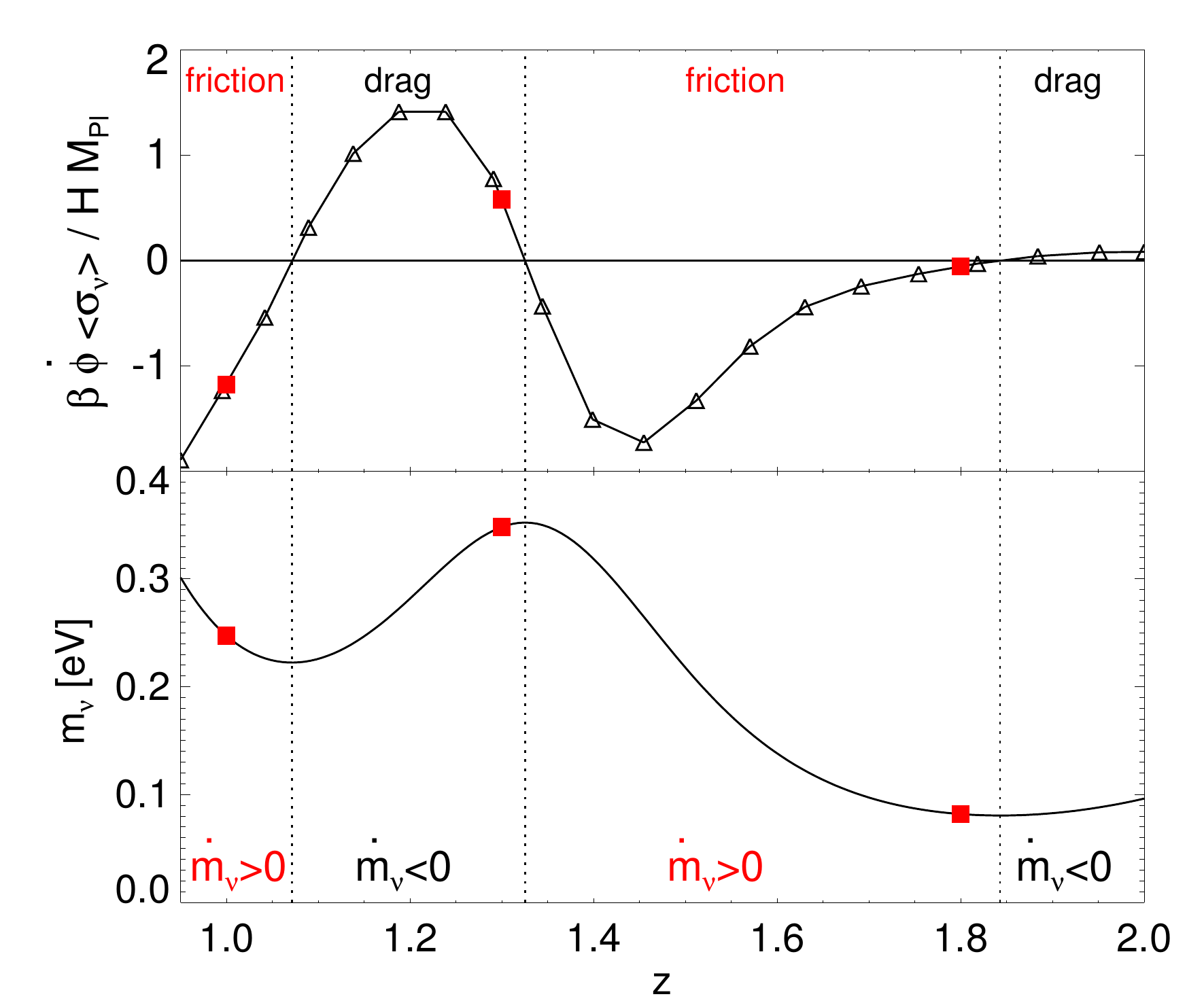}
\caption{The evolution of the average specific friction term $\beta \dot{\phi } <\sigma_{\nu }>/H{M_{{\rm Pl}}}$ as extracted from the Cnu-b simulation of Table~\ref{tab:simulations} ({\em top panel}), and the evolution of the neutrino mass in eV ({\em bottom panel}) as a function of redshift, in the redshift range $0.9 - 2$. The red squares indicate the values of the two quantities at the redshifts represented in the different columns of Figs.~\ref{fig:boxes1}, \ref{fig:boxes2}, and \ref{fig:boxes3}, \ie $z=1.8$, $z=1.3$ and $z=1$, respectively.}
\label{fig:friction}
\end{figure}

This behavior can be understood by considering again the evolution of the neutrino mass
and the corresponding oscillatory behavior of the average neutrino velocity discussed in Sec.~\ref{neutrino_vel}. In particular, the 
simultaneous time evolution of the neutrino mass and of the friction term appearing in the neutrino acceleration equation (Eq.~\ref{acceleration})
can account for the results of our simulations. To better understand the situation, one can have a look at Fig.~\ref{fig:friction}, where the 
average friction term $\beta \dot{\phi } <\sigma_{\nu }>/HM_{{\rm Pl}}$ as extracted from our numerical simulations (top panel) and the neutrino mass evolution (bottom panel)
are plotted as a function of redshift in the redshift interval $0.95-2$. The filled red squares indicate the values of the
two quantities in correspondence of the redshifts shown in Fig.~\ref{fig:boxes1}, Fig.~\ref{fig:boxes2} and Fig.~\ref{fig:boxes3}.

As a consequence of the oscillations of the cosmon $\phi $ around the minimum of its effective potential, the friction term $\beta \dot{\phi }v$ also oscillates and changes sign during the evolution of the universe. This implies
that the friction term actually acts as a drag term, \ie accelerates particles in the direction of their motion, whenever positive, while it behaves as
a real friction, \ie decelerates particles, when it is negative. The vertical dotted lines in Fig.~\ref{fig:friction} indicate the locations where
the friction term changes sign. These points of sign inversion clearly correspond to maxima and minima of the neutrino mass 
evolution: whenever the neutrino mass grows in time ($\dot{m}_{\nu } > 0$), the friction term is negative and acts as a real friction on neutrino particles, decelerating their motion,
while when the neutrino mass decreases in time ($\dot{m}_{\nu } < 0$), the friction term becomes positive, acting as a drag term and accelerating particles.
By having in mind the contextual evolution of the neutrino mass and of the friction term, it is possible to account for the results shown in
Fig.~\ref{fig:boxes1}, Fig.~\ref{fig:boxes2} and Fig.~\ref{fig:boxes3}. 

Neutrinos evolve from a highly homogeneous distribution at $z=z_{{\rm nr}}$ and start falling into the CDM potential wells as soon as they become non-relativistic.
Therefore, the first neutrino structures follow the main filamentary pattern of the already evolved CDM distribution, as one can see at $z=1.8$.
However, as soon as some level of inhomogeneity has developed in the neutrino distribution, the scalar fifth-force acting only between neutrino particles
overcomes standard gravity and becomes the main driver of the neutrino evolution.
At this stage, \ie between $z=1.8$ and $z=1.3$ the neutrino mass is constantly increasing, making the source term for the neutrino acceleration in Eq.~\ref{acceleration}
also increasing. In other words, the neutrino potential wells become deeper and accelerate the instability process. At the same time, the friction
term is negative, thereby acting as a real friction, \ie decelerating particles. This reduces the neutrino kinetic energy thereby contributing to trap neutrino particles
into gravitationally bound structures. This is what we can see in our snapshots at $z=1.3$. After this time, however, the neutrino mass starts decreasing
and consequently the neutrino potential wells start decaying, increasing the total energy of bound neutrino lumps. Contextually, the friction term has become positive
and acts as a dragging force, accelerating particles in the direction of their motion, and therefore increasing the neutrino kinetic energy. These two 
phenomena contribute to increase the total energy of a large number of neutrino particles above zero, and to unbind them from the previously collapsed lumps,
releasing a significant fraction of free particles that fill the space between the surviving neutrino halos. This is the situation that we observe in our snapshots 
at $z=1$, where a significant diffuse neutrino component is present again in our simulation boxes.\\

In conclusion, we have shown here for the first time that the oscillations of the scalar field $\phi $ around the minimum of its effective potential
in Growing Neutrino cosmologies have a strong impact on the growth of linear and nonlinear neutrino structures, and determine a related oscillatory
behavior of the evolution of neutrino density fluctuations. In other words, neutrino structures that form in the context of Growing Neutrino models do not
continuously grow, but rather exhibit a sort of ``pulsation" due to the oscillations of the DE scalar field that drives the growth of the neutrino mass.
As a consequence of this ``pulsation", also the gravitational potential associated to neutrino lumps and large scale structures will not evolve 
monotonically, but rather follow an oscillatory behavior, with the time derivative $\dot{\Phi }_{\nu }$ of the gravitational potential induced by neutrino 
overdensities changing sign along with the oscillations of the cosmon $\phi $.
This non-monotonic evolution of the gravitational potential time derivative $\dot{\Phi }_{\nu }$ could significantly reduce the impact of Growing Neutrino models
on the amplitude of the ISW effect for the low-multipole CMB angular power spectrum with respect to previous linear and nonlinear estimates \citep[as \eg][]{Pettorino_etal_2010}.
A quantification of the ISW effect associated to the neutrino halo ``pulsation" will therefore be necessary in order to test the viability of the model and could
provide a direct observational way to constrain the neutrino coupling. A non-monotonic evolution of the neutrino-induced gravitational potential could also alleviate the tension between observed bulk velocities and the ISW-correlations between temperature fluctuations and observed structures \citep{Ayaita_etal_2009}.

\subsection{Neutrino halo mass function} \label{subsec:statistics}

As a next step in our analysis, we aim to quantify the number density of neutrino lumps as a function of their total mass
and investigate how this number density evolves in time. We are therefore seeking to extract from our numerical simulations
a ``neutrino halo mass function", analogous to the standard CDM halo mass function. To this end, we first need to build a catalog of neutrino structures
formed in our simulated cosmologies. We therefore identify neutrino groups with a Friends-of-Friends algorithm with linking length $\lambda = 0.2\times \bar{d}$
where $\bar{d}$ is the mean interparticle spacing, and we consider only groups containing at least $200$ particles.
For each of these groups we compute the total group mass $M_{{\rm FoF}}$ and the position of the center of mass. 

\begin{figure}
\includegraphics[scale=0.4]{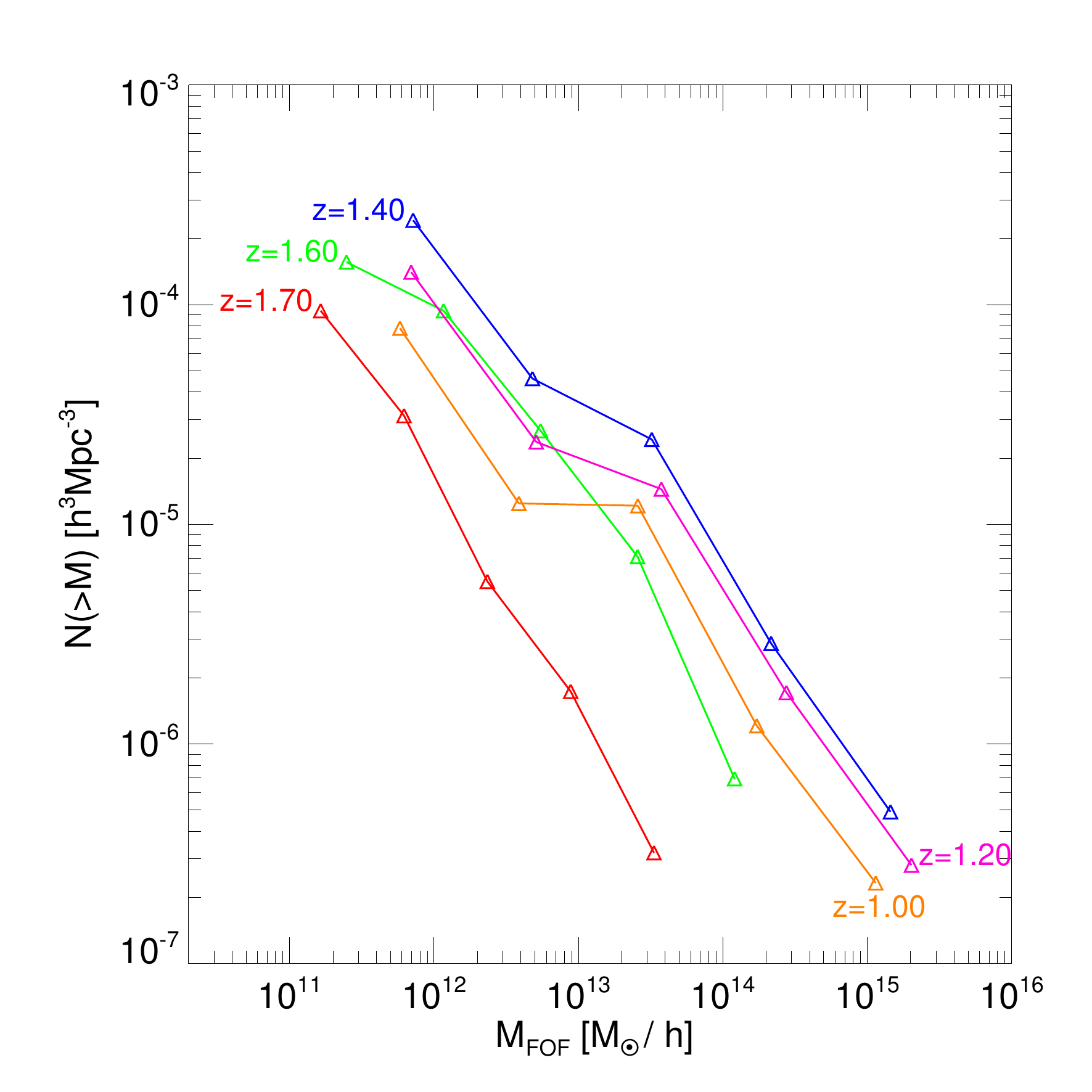}
\caption{The cumulative neutrino halo mass function for the full sample of neutrino halos obtained by combining the separate neutrino halo catalogs of the three simulations Cnu-a, Cnu-b, and Cnu-c. Different curves show the neutrino halo mass function at different redshifts from $z=1.7$ (red) to $z=1$ (orange). The oscillation of neutrino structures is clearly visible in the inversion of trend for the neutrino halo mass function before and after $z=1.4$.}
\label{fig:massfunction}
\end{figure}

We compute our catalogs at different redshifts for the three Growing Neutrino simulations Cnu-a, Cnu-b, and Cnu-c, and we then
combine them into a single neutrino halo mass function in order to maximally extend its mass	range. To do so, we build a single
catalog of all the halos identified in the three different simulations and we associate to each halo the corresponding number density
computed from the mass function extracted from its original simulation. 
This new halo catalog is then ordered by halo mass, and the halos are binned in five logarithmic
mass bins covering the whole mass range of the catalog. The number density in each mass bin is then computed by averaging 
the original number density of all the bin members. The result of this analysis is shown in Fig.~\ref{fig:massfunction}, where the full
neutrino halo mass function extracted from the combination of the three simulations Cnu-a, Cnu-b, and Cnu-c is plotted
for the five mass bins of our full catalog and for five different redshifts. This result represents the first estimate ever made for the abundance of neutrino
lumps in a realistic realization of the nonlinear evolution of a Growing Neutrino cosmology.

Once more, as discovered by the visual inspection of the neutrino distribution as a function of redshift discussed in Sec.~\ref{lss}, 
we can notice that the abundance of neutrino lumps does not grow monotonically in time. On the contrary, as Fig.~\ref{fig:massfunction}
clearly shows, the number density of neutrino lumps at all masses in the mass range covered by our halo sample continuously grows 
from $z=1.8$ until $z=1.4$ (see the red, green, and blue curves in Fig.~\ref{fig:massfunction} representing $z=1.7$, $z=1.6$, and $z=1.4$, respectively)
reaching a maximum amplitude at $z=1.4$. This is consistent with the interpretation given in Sec.~\ref{lss} and exemplified in Fig.~\ref{fig:friction}:
between $z=1.8$ and $z=1.4$ the neutrino mass grows in time making the neutrino potential wells progressively deeper, and at the same time the
friction term acts as a real friction, slowing down neutrino particles and favoring the formation of bound structures, whose number density
consistently grows in time. 

At $z=1.4$, however, the scalar field $\phi $ inverts its motion along its oscillations around the minimum of its effective potential,
and consequently the neutrino mass starts decreasing, inducing the decay of the already formed neutrino potential wells. As a consequence of the 
inversion of the scalar field motion, also the friction term changes its sign, becoming an effective drag that accelerates the motion of neutrino particles,
thereby increasing the neutrino kinetic energy (see Fig.~\ref{fig:friction}). This contributes to unbind a significant number of neutrino particles from previously bound structures,
and determines the ``evaporation" of a fraction of the neutrino lumps that had formed in the previous stage of neutrino mass growth. 
This phenomenon can be clearly identified in Fig.~\ref{fig:massfunction}, where the purple and the orange curves, representing the neutrino 
halo mass function at $z=1.2$ and $z=1$, respectively, show a progressively smaller number density of neutrino lumps at all masses
with respect to the maximum amplitude of the neutrino mass function reached at $z=1.4$.

Once again, we are witnessing an oscillatory behavior of the growth of large scale neutrino structures.
Our results therefore show for the first time that neutrino structures forming in the context of Growing Neutrino cosmologies are not stable objects
but rather experience subsequent phases of growth and decay, driven by the background evolution of the cosmon.
We call this behavior ``pulsation" of the large scale neutrino structures.

The detailed investigation of the ``pulsation" of neutrino lumps and of their associated gravitational potential, 
combined with an estimate of the abundance of neutrino structures as a function of mass at different redshifts as provided by Fig.~\ref{fig:massfunction}, 
will allow a determination of the evolution of the overall cosmological gravitational potential in Growing Neutrino cosmologies.

\subsection{Drag of Cold Dark Matter structures} \label{subsec:cdm}
\begin{figure*}
\includegraphics[scale=0.4]{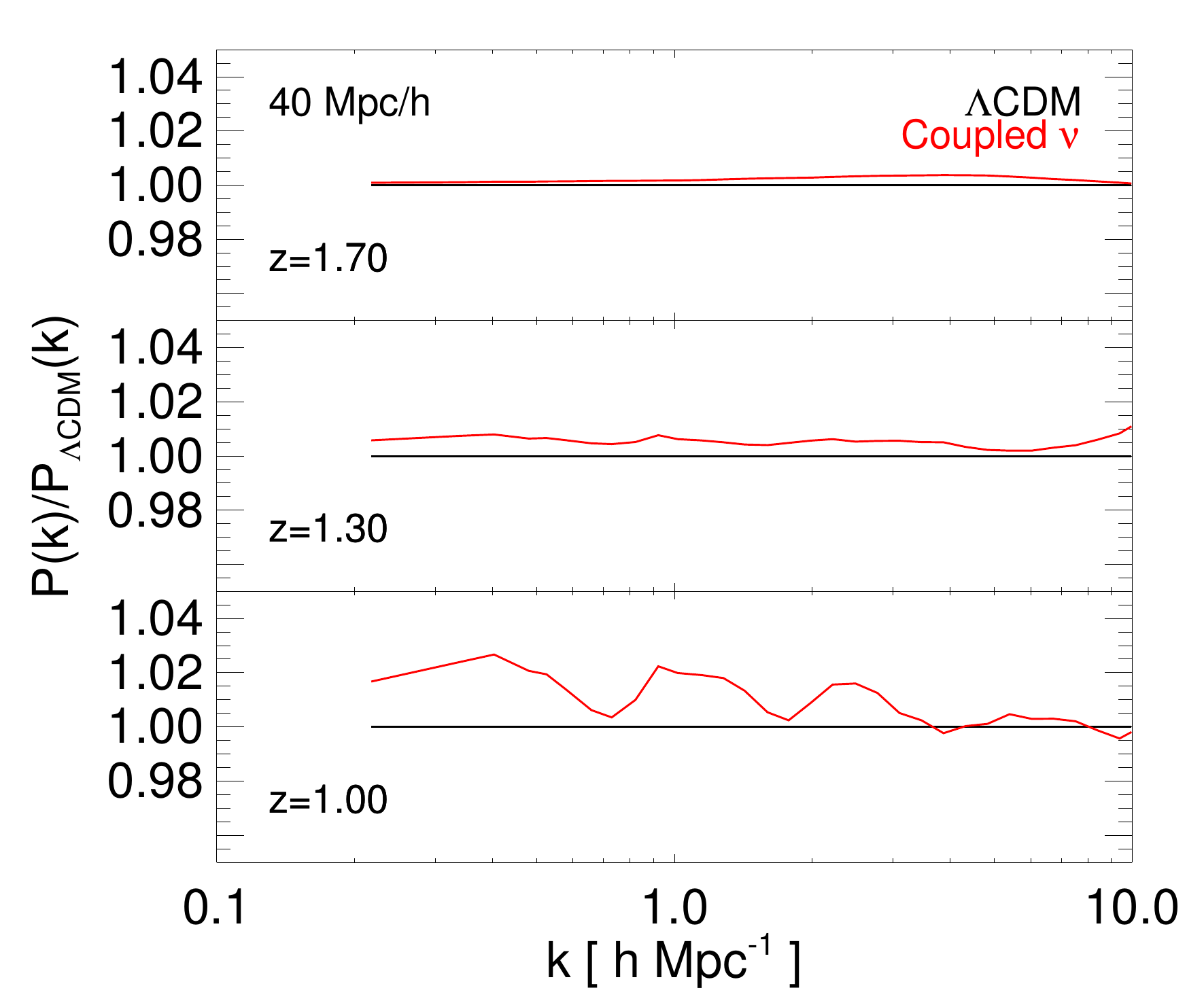}
\includegraphics[scale=0.4]{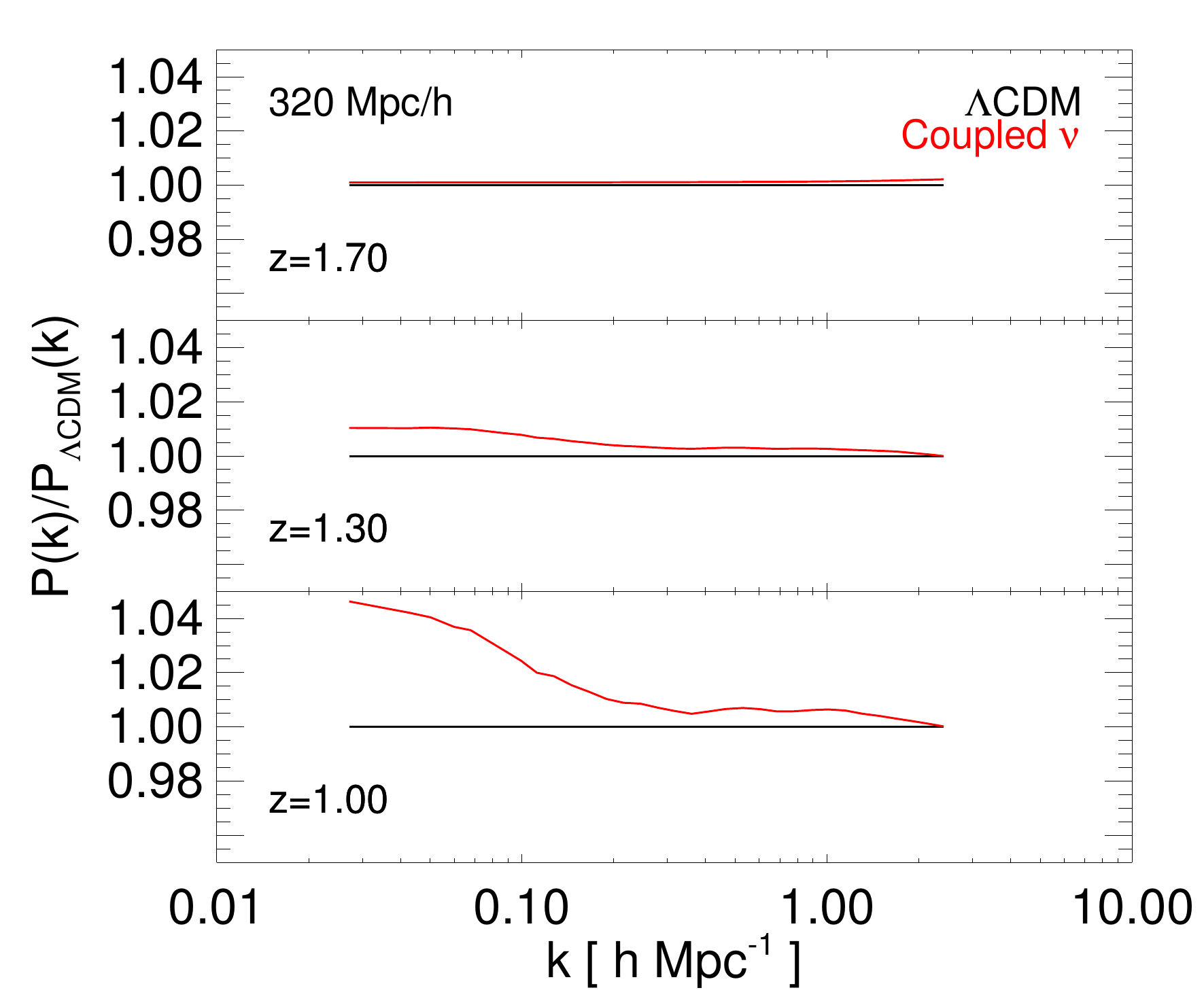}
\caption{Ratio of the CDM power spectrum in Growing Neutrino simulations to
 standard $\Lambda $CDM simulations at different redshifts for the a ({\em left}) 
and c ({\em right}) simulations of Table~\ref{tab:simulations}.}
\label{fig:powerspectra}
\end{figure*}

As a final step in our analysis, we now try to estimate the impact that the coupling between the scalar field $\phi $
and massive neutrinos can have on the evolution of CDM large scale structures. Since no coupling between DE and CDM
is present in the specific scenario under investigation in the present work, the only way in which our model can alter
the evolution of structure formation for the CDM density field is through a backreaction effect of the large neutrino structures
that form below $z_{{\rm nr}}$ onto the dynamics of CDM particles. In particular, the formation of large neutrino structures
at high redshifts can generate significant gravitational potential wells on large scales that might have a measurable impact on the
properties of CDM large scale structures.

In order to quantify this impact, we will make use of the three $\Lambda $CDM simulations described in Table~\ref{tab:simulations}
to directly compare the evolution of CDM structures in the presence of a coupling between DE and massive neutrinos
to the standard $\Lambda $CDM evolution.

\subsubsection{The CDM power spectrum}

First of all, we compute the CDM density power spectrum $P(k)$ as a function of wavenumber $k$ in all our simulations, and
we compare the results obtained from simulations of the same scale with and without a coupling between DE and massive neutrinos.
The results are shown in Fig.~\ref{fig:powerspectra}, where the ratio $P(k)/P_{\Lambda {\rm CDM}}(k)$ between
the CDM power spectrum computed with and without neutrino coupling is shown for three different redshift values, namely
$z=1.7$, $z=1.3$, and $z=1$, for the smallest (left panel) and the biggest (right panel) box in our simulations set.

As Fig.~\ref{fig:powerspectra} shows, at $z=1.7$ there is basically no difference between the two power spectra in any of the shown simulation pairs.
This indicates that at this early time, the low neutrino mass and the small number of neutrino structures present in the universe are not
sufficient to influence the dynamics of CDM structures, and that at this stage CDM is still acting as a seed for the development of the first neutrino overdensities.

At $z=1.3$, instead, an increase of CDM power of the order of $\sim 1 \%$ at all scales in the simulations of Growing Neutrino cosmologies as compared to $\Lambda $CDM
appears clearly in both plots. The largest scale simulations ($\Lambda $CDM-c and Cnu-c, right panel of Fig.~\ref{fig:powerspectra}) 
also show a weak scale dependence of the power enhancement, that appears to be larger at the largest scales of the simulation. This shows that at $z=1.3$
the gravitational potential induced by the formation of large neutrino structures has already a small but non-negligible impact on the underlying CDM distribution.
The roles of these two cosmic component are therefore inverted: while at earlier times the CDM drives the motion of massive neutrino particles, acting as a seed for the
formation of the first neutrino lumps, now it is the fast growth of neutrino structures to mildly influence the evolution of the CDM distribution on very large length scales.

The effect becomes more pronounced at $z=1$, where the increase in CDM power can reach a level of $3-4\%$ for wave numbers smaller than $k\sim 0.1$ h/Mpc (left 
panel of Fig.~\ref{fig:powerspectra}). At these scales, the ratio of power spectra shows significant irregularities. These are due to the statistical limitation
determined by the box size, where only a few neutrino lumps are still present at $z=1$ in the $40$ (as well as in the $120$ Mpc/h) simulation boxe (see Fig.~\ref{fig:boxes1}). This clearly induces an
excess of power for some specific scales (determined by the specific spatial distribution of the lumps)
that is reflected in the irregular shape of the power spectrum ratio in the left panel of Fig.~\ref{fig:powerspectra}.
Our largest coupled neutrino simulation, however, with a size of $320$ Mpc/h still contains a significant number of neutrino lumps at $z=1$
(see Fig.~\ref{fig:boxes3}), thereby providing a statistically more reliable estimate of the CDM power spectrum. For this simulation (right panel of Fig.~\ref{fig:powerspectra}),
in fact, we see that no significant irregularity is present in the ratio of the power spectra with and without growing neutrinos, and we can more easily 
quantify the backreaction of neutrino structures on CDM. As the plot shows, there is a clear scale dependence of the power enhancement, that reaches values of the order
of $\sim 5\%$ at the largest scales of the simulation, while never exceeding $2\%$ for scales smaller than $k\sim 0.1$ h/Mpc.

The backreaction effect of the formation of neutrino halos on the CDM large scale structures is therefore a potentially observable feature of the model,
especially at very large scales, and could contribute to significantly constrain the Growing Neutrino scenario.

At this point we also mention another characteristic feature of structure formation in Growing Neutrino quintessence. Measurements of the power spectrum by distributions of galaxies presumably give access to the CDM power spectrum, if we assume that baryons trace CDM (up to bias). On the other hand, for a measurement of the power spectrum by weak lensing the propagation of photons is influenced by the total gravitational potential, including the one induced by neutrino lumps. The power spectrum ``seen'' by photons is a spectrum for fluctuations of combined CDM, baryons and neutrinos. It may differ from the CDM spectrum probed by galaxy distributions, leading to inconsistencies between the observed galaxy and lensing power spectra if a standard $\Lambda $CDM cosmology is assumed as the true underlying model.

\subsubsection{The CDM bulk flow}
\begin{figure*}
\includegraphics[scale=0.28]{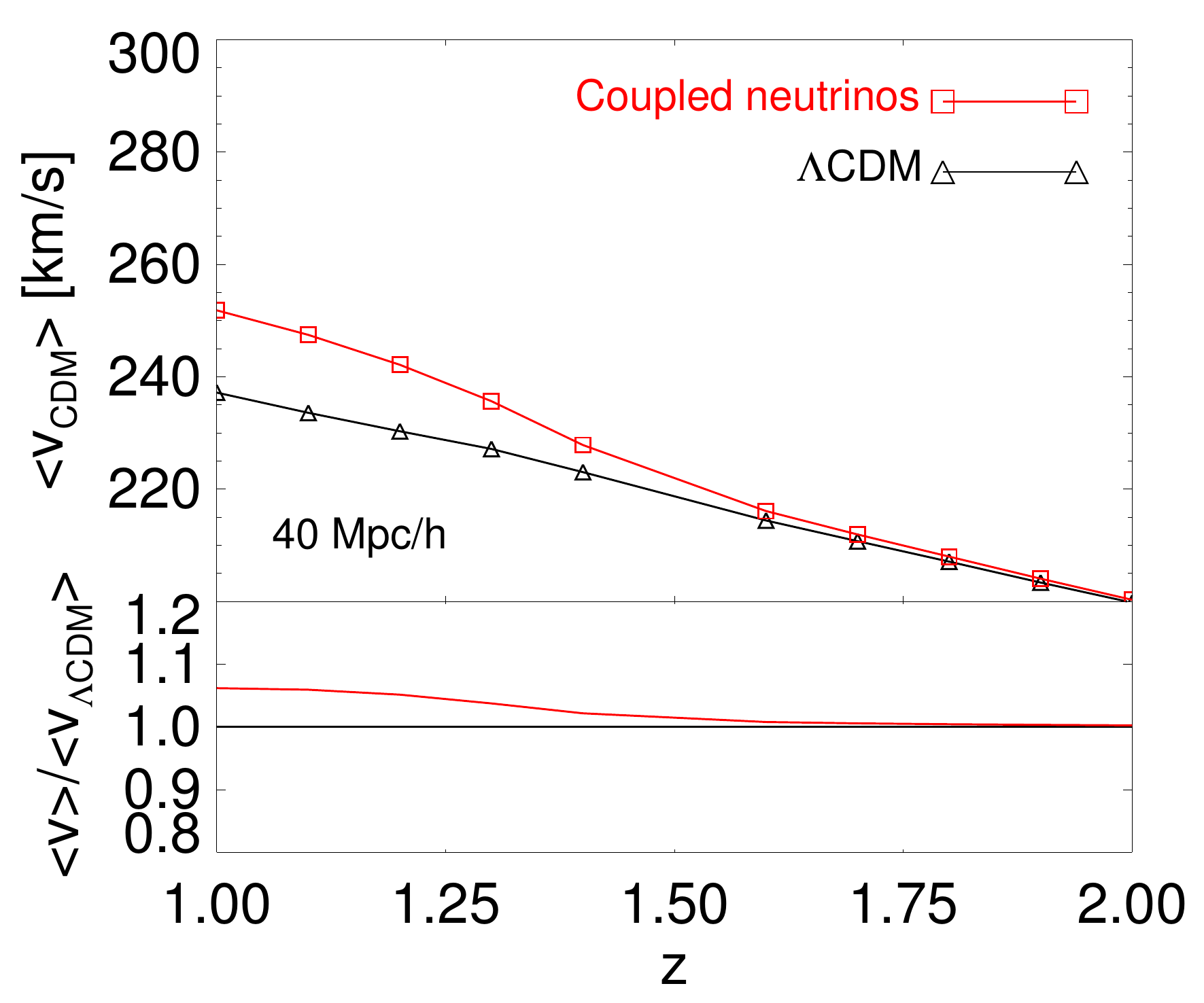}
\includegraphics[scale=0.28]{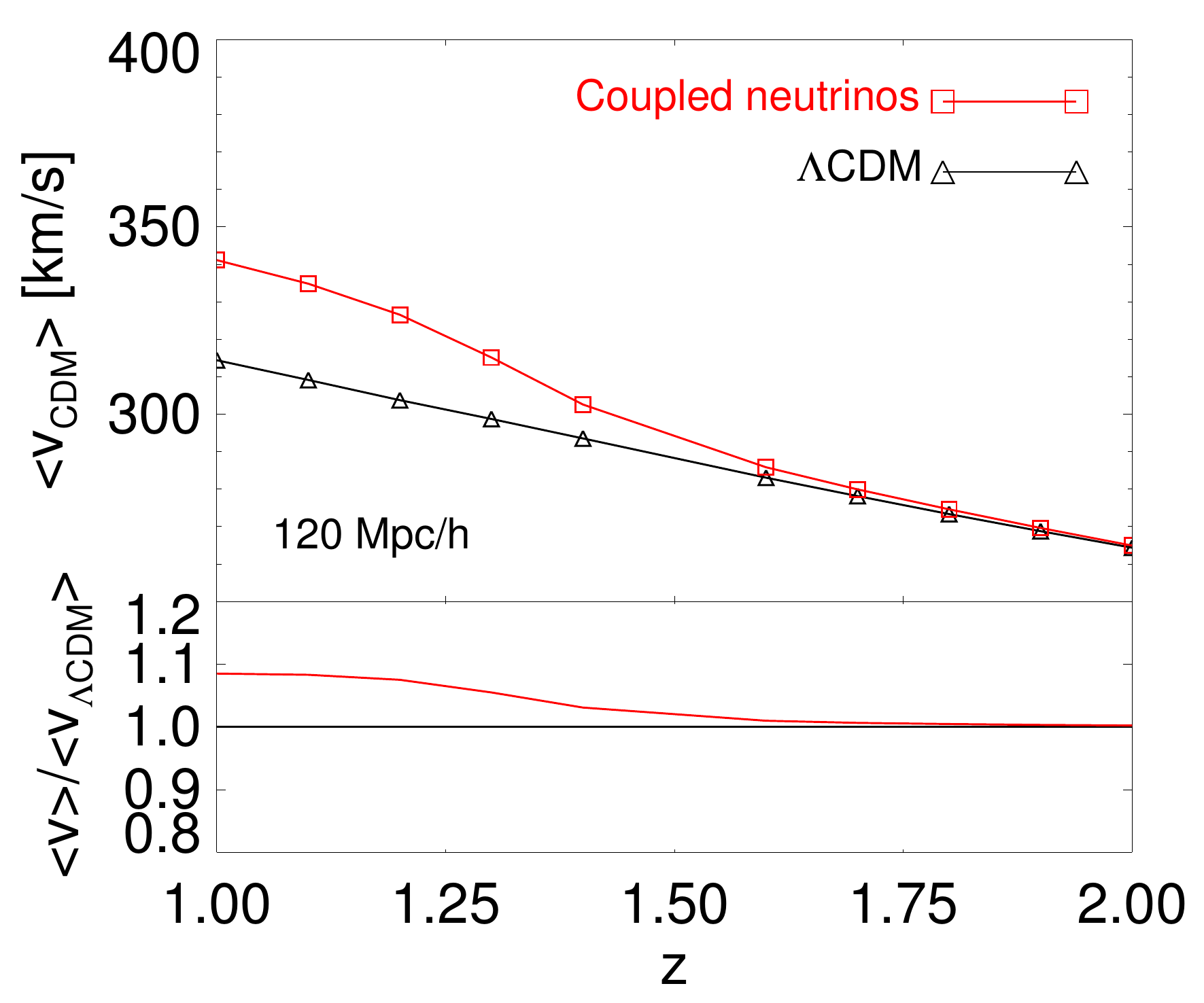}
\includegraphics[scale=0.28]{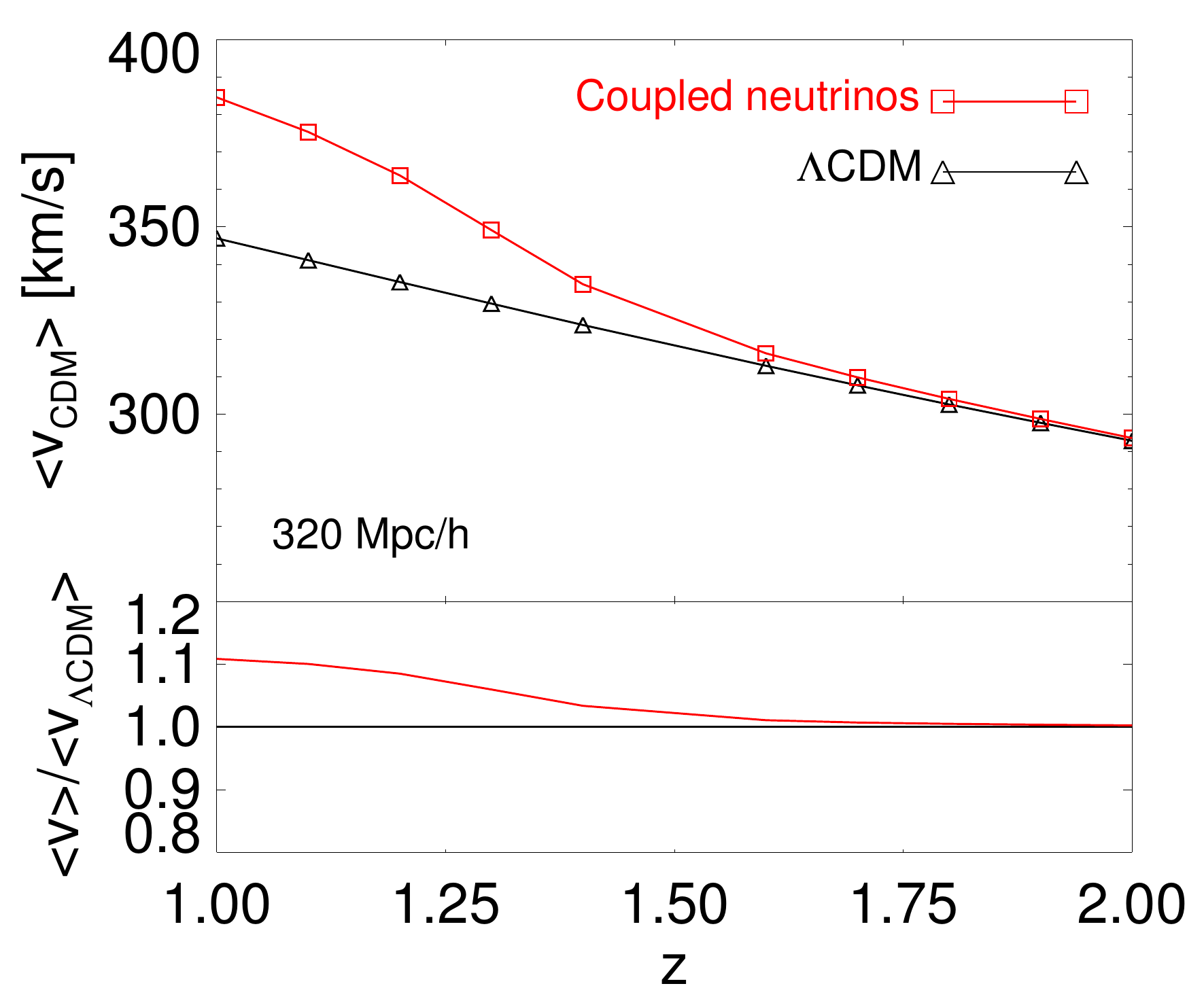}
\caption{The average CDM peculiar velocity ({\em top panels}) in Growing Neutrino simulations (red squares) and in the corresponding standard $\Lambda $CDM simulations (black triangles) as a function of redshift for the a ({\em left}), b ({\em middle}) and c ({\em right}) simulations of Table~\ref{tab:simulations}. The {\em bottom panels} show the ratio of the average CDM peculiar velocities in the two different cosmological scenarios. An enhancement of CDM velocities at low redshifts in the Growing Neutrino models as compared to $\Lambda $CDM is clearly visible in all the different simulations.}
\label{fig:bulk_flows}
\end{figure*}

As a last step in our analysis we consider the impact that the growth of large scale neutrino structures can have on CDM peculiar velocities.
We compute the average CDM peculiar velocity for all our simulation snapshots and we compare its time evolution for the coupled neutrino
simulations with the corresponding $\Lambda $CDM runs. The results of this analysis are shown in the three plots of Fig.~\ref{fig:bulk_flows} for
the three different box sizes of our simulations, namely $40$, $120$, and $320$ Mpc/h in the left, center, and right plots, respectively.

All the three plots show a clear increase of the average CDM peculiar velocity at redshifts where the large scale neutrino structures start
playing a role in the coupled neutrino cosmologies. As a consequence of the sudden development of very large scale gravitational potential wells
determined by the growth of neutrino lumps, CDM particles receive an additional acceleration with respect to the standard $\Lambda $CDM scenario,
and progressively acquire larger velocities. This effect, as shown in Fig.~\ref{fig:bulk_flows}, determines an increase of the average CDM velocity
of the order of $5-10\%$, and shows a clear scale dependence, with the weakest impact ($\sim 5\%$) for the $40$ Mpc/h simulation, where only 
a few mid-size ($5-10$ Mpc/h) neutrino lumps are present at the redshifts under consideration, and the largest impact ($\sim 10\%$) for the 
$320$ Mpc/h simulation, where instead a large number of intermediate-to-large size ($10-50$ Mpc/h) neutrino lumps can source the additional gravitational potential
responsible for the velocity excess.

We have therefore shown that an anomalously large CDM velocity field might indicate the presence of an unexpected large-scale gravitational potential
and that such potential could be sourced in our model by large scale neutrino structures.
Growing neutrino cosmolgies might therefore provide a natural explanation to the claimed detected large CDM bulk flow on scales larger
than $\sim 100$ Mpc/h \citep{Watkins_etal_2009}. 

Our preliminary analysis involves only the average peculiar velocity of all the CDM
particles in the simulations. For a direct comparison with observational data, however, a determination of the velocity field for visible tracers of the CDM
distribution through the construction of a full mock halo catalog would be required. We defer this analysis to a follow up paper.

It is interesting to notice that the growth of the neutrino gravitational potential at $z > 1.3$ is sufficient to induce a $10\%$ velocity excess with respect to 
$\Lambda $CDM on scales larger than $~300$ Mpc/h, despite the subsequent decay of the neutrino gravitational potential at $1.0 < z < 1.3$.
This clearly shows that a detected anomalous CDM bulk flow in the low-redshift universe might not imply the simultaneous existence of anomalously large-scale
gravitational potentials, but could be explained, as for the case of the Growing Neutrino models investigated in this work, by an epoch of subsequent growth and 
decay  of such potentials at high redshifts ($z > 1$) whose effects remain imprinted in the CDM velocity distribution. Indeed, if only gravity provides for the attraction of particles, the bulk flow and the density contrast are tightly correlated. Enhanced bulk flow combined with only a mild increase in the density contrast would require additional attraction beyond gravity \citep{Ayaita_etal_2009}, as realised in our model. 

As pointed out above, such a peculiar time evolution (that we defined as a ``pulsation") of the large-scale gravitational potential 
might imprint distinctive features on the large-scale CMB angular
power spectrum through the ISW effect, and a detailed investigation of these possible features, which goes beyond the scope of this paper, will be a necessary 
test in order to assess the viability of the model. 

\section{Conclusions}
\label{concl}

In the present paper, we have described the results of the first N-body simulations of structure formation in the context of Growing Neutrino cosmologies. Our numerical treatment allowed for the first time to follow the fully nonlinear evolution of a statistically meaningful sample of neutrino lumps forming in a specific Growing Neutrino model up to $z=1$. This limiting redshift is due to the rapid growth of neutrino velocities into the relativistic regime, which does not fulfill the condition of small velocities that defines the newtonian limit for which standard N-body algorithms are implemented. Substantial modifications of the presently available numerical codes will be required in order to include in the treatment relativistic corrections, so to extend the reliability of numerical simulations down to $z=0$. Nevertheless, our results significantly improve previous analyses by following the nonlinear evolution of neutrino lumps to lower redshifts and well beyond the onset of virialisation. Furthermore, the use of N-body simulations allows to include the effects of directionality in the neutrino drag force and of merging and interaction of a realistic distribution of neutrino lumps.

Our analysis has highlighted for the first time the phenomenon of pulsation in the growth of the density contrast in the neutrino fluid. In all realistic models of Growing Neutrino quintessence an oscillatory change of the neutrino mass has been found to set in once the interaction with the neutrino fluid becomes important for the time evolution of the cosmon and therefore for dark energy. An oscillatory neutrino mass results in a periodic change between friction and drag for the neutrino velocities. For periods of increasing neutrino mass an effective friction term accounts for the fact that for constant momentum the velocity is reduced due to the increase in mass. In contrast, for periods of decreasing neutrino mass the opposite ``drag term'' enhances the neutrino velocities. Furthermore, the strength of the attractive force between neutrinos is proportional to the squared neutrino mass and shows oscillatory behavior for oscillatory mass. The basic ingredients for the pulsation, i.e. the subsequent increase and decrease in the neutrino velocities and density contrast that we observe in our N-body simulations, seem to be rather robust. We therefore expect that the phenomenon of pulsation will be found also in more accurate treatments of the nonlinear growth of density fluctuations, which may overcome the limitations of the present numerical investigation.

A pulsation in the growth of structure does not occur in the standard $\Lambda$CDM scenario. This phenomenon could therefore lead to interesting tests of the Growing Neutrino model. In particular, the CDM bulk flow may be enhanced despite very moderate modifications of the density contrast as compared to the $\Lambda$CDM model. Also the tension between the size of observed correlations of the ISW-effect in CMB anisotropies with observed structures on one side, and the observed bulk flow on the other side, could possibly be reduced. Quantitative statements will need, however, a treatment of the nonlinear structure growth that goes beyond the limitations of the N-body simulations presented in this work.

A second important observation of the present investigation concerns the overall strength of the neutrino induced gravitational potential. We defer the quantitative analysis of the fluctuations in the gravitational potential to a separate publication. Nevertheless, already at the present stage it becomes apparent that the gravitational potential does never become very large for the redshifts $z>1$ and the size of the boxes investigated in this paper. This can be inferred from the very modest increase in the CDM power spectrum shown in Fig.\ref{fig:powerspectra}. This qualitative finding of limitations in the growth of the fluctuations in the gravitational potential confirms the arguments advocated by \cite{Pettorino_etal_2010}. It clearly demonstrates that an extrapolation of the linear growth of neutrino density fluctuations and the neutrino induced gravitational potential \citep[see \eg ][]{Franca_etal_2009} leads to huge overestimates for these quantities.

Our third finding concerns the large neutrino velocities that are induced by the strong mutual attraction of lumps due to the cosmon force. The observation in our simulations that neutrino velocities become comparable to the velocity of light, limits the range of validity of our treatment. On the other hand, the effective pressure for the neutrino fluid that is induced by these velocities counteracts the strong attraction mediated by the cosmon force. This pressure will limit the formation of very large and deep gravitational potential wells induced by neutrino lumps. In turn, the quantitative understanding of the evolution of fluctuations in the gravitational potential is crucial for the ISW-effect in CMB anisotropies. A quantitative computation of the imprint of Growing Neutrino quintessence for the CMB will have to wait for a quantitative computation of the cosmological neutrino induced gravitational potential.

Presumably, this will require an extension of our methods to include effects from relativistic velocities.
We do not know yet to what extent such effects will reduce the strength of the pulsation in the range $z\le 1$.

Finally, a fundamental new result of the present work is the determination of the statistical distribution of neutrino halos as a function of redshift.
For the first time we have attempted a computation of the neutrino halo mass function and of the quantitative effects on the CDM power spectrum and bulk flow.
These estimates are still limited in redshift range to $z\le 1$, and substantial uncertainties remain from the validity of the use of non-relativistic equations and from neglecting
``backreaction" effects as well as locally varying neutrino masses.
Nevertheless, interesting effects which may lead to observational tests for the model become visible. These concern, in particular, an enhanced CDM bulk flow and the
presence of very large scale neutrino lumps and associated gravitational potential wells.

\section*{Acknowledgments}

This work has been supported by 
the TRR33 Transregio Collaborative Research
Network on the ``Dark Universe'', and by 
the DFG Cluster of Excellence ``Origin and Structure of the Universe''.
Support was given to V.P. by the Italian Space Agency through the ASI
contracts Euclid-IC (I/031/10/0).
All the numerical simulations have been performed on the Linux Cluster at the LRZ computing centre in Garching.

\section*{Appendix}

In this paper we have applied the standard Newtonian equation of motion
to all particles in the simulation. However, as we have seen, neutrinos
reach soon relativistic velocities and after $z\approx1$ the assumption
of non-relativistic motion is no longer acceptable. For completeness
and in view of future applications we write down the equations of
motion in the weak and slow-varying field limit but without limitations
on velocities. We also include a non-linear scalar field potential.

We assume the usual longitudinal metric for scalar perturbations \begin{equation}
ds^{2}=-(1+2\Psi)dt^{2}+a^{2}(1+2\Phi)dx^{i}dx_{i}\end{equation}
Defining the affine parameter $\tau$ such that $d\tau^{2}=-ds^{2}$
we can write\begin{equation}
d\tau=dt(1+\Psi)\Gamma\end{equation}
where for small $\Phi,\Psi$,\[
\Gamma\equiv[1-v^{2}(1+2(\Phi-\Psi))]^{1/2}\]
and where we define the peculiar velocity \begin{equation}
v^{i}\equiv a\frac{dx^{i}}{dt}\end{equation}
We need to derive the geodesic equations for a particle\begin{equation}
\frac{du^{\mu}}{d\tau}+\Gamma_{\alpha\beta}^{\mu}u^{\alpha}u^{\beta}=0\end{equation}
where the four velocity is\begin{equation}
u^{\alpha}=\frac{dx^{\alpha}}{d\tau}=\{\frac{1-\Psi}{\Gamma},\frac{v^{i}}{a\Gamma}\}\end{equation}
Here we assume that the gravitational potentials are small, but the
velocity can be relativistic. The geodesic equation holds for an uncoupled
particle; below we extend it to coupled particles. As usual in the
weak-field limit, we assume from now on that the time derivatives
of $\Psi,\Phi$ are negligible. 
In order to keep the equations linear in the potentials, we
need to assume also $\gamma^2 \Psi \ll 1$ (and similarly for
$\Phi,\varphi$).
Notice that $d/dt$ is a total derivative
and therefore \begin{equation}
\frac{d(v^{i}(1-\Psi)/\Gamma)}{dt}\approx\frac{1}{\Gamma}(\dot{v}^{i}+v^{i}\frac{v\dot{v}}{\Gamma^{2}}-\frac{v^{i}v^{j}\Psi_{,j}-v^{2}v^{i}v^{j}\Phi_{,j}}{a\Gamma^{2}})\end{equation}
where $v=|\mathbf{v}|$ .

From the geodesic equation with $\mu=i$ we obtain to first order
in $\Phi,\Psi$:
\begin{eqnarray}
& &a\dot{v_{i}}(1+\varepsilon_{1})+\gamma^{2}av_{i}v\dot{v}(1+\varepsilon_{2}) = \nonumber \\
& &-aHv_{i}(1+\varepsilon_{1}) + v^{2}\Phi_{,i} - (\gamma^{2}+1)v_{i}(v_{j}\nabla_{j}\Phi) + \nonumber \\
& &\gamma^{2}v_{i}(v_{j}\nabla_{j}\Psi)-(\Psi_{,i}-v^{2}\Phi_{,i})
\end{eqnarray}
with\begin{equation}
\varepsilon_{n}=2\Psi(1-2n\gamma^{2})\end{equation}
and where we used the usual special-relativistic factor \begin{equation}
\gamma^{2}=(1-v^{2})^{-1}\end{equation}
Now as long as dark matter density is the dominating component, we
can assume that the anisotropic stress induced by the relativistic
neutrinos is negligible and therefore we put $\Phi=-\Psi$. Then we
obtain
\begin{eqnarray}
& &\dot{v_{i}}(1+\varepsilon_{1})+\gamma^{2}v_{i}v\dot{v}(1+\varepsilon_{2}) = \nonumber \\
& &-Hv_{i}(1+\varepsilon_{1})+(2\gamma^{2}+1)v_{i}(v_{j}\nabla_{j}\Psi)-(1+v^{2})\Psi_{,i}\label{eq:spatialnocoup}
\end{eqnarray}
and from now on all spatial derivatives are with respect to physical
distances $r_{i}=ax_{i}$. The $\mu=0$ equation is\begin{equation}
\gamma^{2}v\dot{v}(1+\varepsilon_{2})=-v^{2}H(1+\varepsilon_{1}-4\Psi)+(2\gamma^{2}-3)(v_{j}\nabla_{j}\Psi)\end{equation}
(this can also be obtained up to terms higher order in $\Psi$ by
multiplying Eq. (\ref{eq:spatialnocoup}) by $v^{i}$) and we can
use it to simplify Eq. (\ref{eq:spatialnocoup}) obtaining \begin{equation}
\gamma^{2}\dot{\mathbf{v}}(1+\varepsilon_{1})=-H\mathbf{v}(1-2\Psi)+4\gamma^{2}\mathbf{v}(\mathbf{v}\cdot\mathbf{\nabla}\Psi)-(1+2v^{2}\gamma^{2})\mathbf{\nabla}\Psi\end{equation}

Now, to generalize to the coupled model, we can derive the coupled
equation by conformally transforming the geodesic equation. If we
put\begin{equation}
\hat{g}_{\mu\nu}=e^{2\beta\phi}g_{\mu\nu}\end{equation}
we obtain\begin{equation}
\Gamma_{\nu\lambda}^{\mu}=\hat{\Gamma}_{\nu\lambda}^{\mu}-\beta(\phi_{,\lambda}\delta_{\nu}^{\mu}+\phi_{,\nu}\delta_{\lambda}^{\mu}-\phi^{,\mu}\hat{g}_{\nu\lambda})\end{equation}
and \begin{equation}
u^{\mu}=e^{\beta\phi}\hat{u}^{\mu}\end{equation}
Then, by assuming also that $\varphi$ is taken at first order and
$\dot{\varphi}$ is negligible we have
\begin{eqnarray}
& &\dot{\mathbf{v}}(1+\varepsilon_{1})+\gamma^{2}\mathbf{v}v\dot{v}(1+\varepsilon_{2}) = \nonumber \\
& &-\tilde{H}\mathbf{v}(1+\varepsilon_{1})-\nabla(\Psi-\beta\varphi)-v^{2}\nabla(\Psi+\beta\varphi)+\nonumber \\
& &\mathbf{v}(\mathbf{v}\cdot\mathbf{\nabla}(\Psi+\beta\varphi))+2\gamma^{2}\mathbf{v}(\mathbf{v}\cdot\nabla\Psi)\label{eq:gvar-1}
\end{eqnarray}
where $\tilde{H} = H(1-\beta\frac{\dot{\phi}}{H})$ . This reduces to
the standard case \begin{equation}
\dot{\mathbf{v}}=-\tilde{H}\mathbf{v}-\nabla(\Psi-\beta\varphi)\label{stand}\end{equation}
for small velocities. The $\mu=0$ equation is\begin{equation}
\gamma^{2}v\dot{v}(1+\varepsilon_{2})=-\tilde{H}v^{2}(1+\varepsilon_{1}-4\Psi)+(\mathbf{v}\cdot\mathbf{\nabla}((2\gamma^{2}-3)\Psi+\beta\varphi))\label{eq:gvar-1-1}\end{equation}
which again can be used to simplify Eq. (\ref{eq:gvar-1}):

\begin{eqnarray}
& &\gamma^{2}\dot{\mathbf{v}}(1+\varepsilon_{1}) = \nonumber \\
& &-\tilde{H}\mathbf{v}(1-2\Psi)+4\gamma^{2}\mathbf{v}(\mathbf{v}\cdot\mathbf{\nabla}\Psi)-\nabla(\Psi-\beta\varphi)-2\gamma^{2}v^{2}\nabla\Psi \nonumber \\
\label{eq:gvar-1-2-2}
\end{eqnarray}
This can be further generalized by including the higher-order terms
in the perturbed field $\varphi$:
\begin{eqnarray}
& &\gamma^{2}\dot{\mathbf{v}}(1+\varepsilon_{1}) = \nonumber \\
& &-\tilde{H}\mathbf{v}(1-2\Psi)+4\gamma^{2}\mathbf{v}(\mathbf{v}\cdot\mathbf{\nabla}\Psi)-\nabla(\Psi-\beta\varphi) - \nonumber \\
& &2\gamma^{2}v^{2}\nabla\Psi+2\beta\Psi\nabla\varphi\label{eq:gvar-1-2-2-1}
\end{eqnarray}

\bibliographystyle{mnras}
\bibliography{baldi_bibliography}

\label{lastpage}

\end{document}